\documentclass[smallextended,numbook]{svjour3}
\usepackage{lmodern}
\journalname{Journal of Statistical Physics}
\journalname{}

\usepackage{ulem}
\usepackage{array}
\usepackage{graphicx,float,amsmath,amssymb,mathrsfs}
\usepackage[english,french]{babel}

\usepackage{url,hyperref}  

\usepackage[usenames,dvipsnames]{color}
\usepackage[T1]{fontenc}
\begin{document}

\catcode`@=11 % If we need private TeX macros
\renewcommand\tableofcontents{%
  \section*{\contentsname}%
  \@starttoc{toc}%
}
\catcode`@=12% '@' is no more a character

\selectlanguage{english}

\title{Critical behavior and crossover scaling in the Light–Heavy model}

\author{Shilpa Prakash \and Mustansir Barma \and Kabir Ramola}

\institute{
S. Prakash
\at 
Tata Institute of Fundamental Research, Hyderabad, 500046, India\\
\email{pshilpa@tifrh.res.in}
\and
M. Barma
\at
Tata Institute of Fundamental Research, Hyderabad, 500046, India\\
\email{barma@tifrh.res.in}
\and
K. Ramola
\at
Tata Institute of Fundamental Research, Hyderabad, 500046, India\\
\email{kramola@tifrh.res.in}
}

\date{\today}

\maketitle

\vspace{-0.5cm}
\begin{abstract}
The Light-Heavy (LH) model involves two species of particles (light and heavy) coupled with a fluctuating surface (described by tilts). The dynamics include the inherent diffusion of the particles (or tilts) as well as the drive provided by the tilts (or particles). When the two are of similar magnitude, the system lies in the unscaled (uLH) regime, while a significantly weaker drive leads to the scaled (sLH) regime. In the unscaled limit, the model exhibits an order–disorder transition characterized by the fluctuation-dominated phase ordering (FDPO). In this state, interestingly the dynamics is driven by multiple modes, giving rise to dynamic clusters. Away from the critical regime the disordered phase retains vestiges of FDPO behavior on length scales smaller than the correlation length. We examine this local FDPO-like behavior by using a scaling function that links the off-critical and critical regimes. We next turn to the scaled model and show that the multi-mode dynamics present in the unscaled regime is replaced by dynamics that is effectively controlled by a single dominant mode in the scaled regime. Concurrently, the two-point correlations change from the $\mathcal{O}(1)$ FDPO form to an anomalous long-range form that decays as $1/\sqrt{L}$. Drawing on the analogy with the sABC model, where similar anomalous correlations appear at criticality, we derive an analytical expression for the two-point correlation function using the same approach used for that model.  
\end{abstract}

%\vspace{0.25cm}

\noindent
{\small
%\textit{PACS numbers}~: 72.15.Rn ; 02.50.-r
}

\tableofcontents 

\section{Introduction}
\label{sec:intro}

Driven diffusive systems~\cite{SCHMITTMANN1998} can display qualitatively distinct large-scale behavior, depending on how the strength of the drive compares with diffusion. If the drive remains of the same order as diffusion in the thermodynamic limit, the dynamics are strongly out of equilibrium and often give rise to rich phase diagrams and collective effects that are challenging to characterize analytically. An alternative perspective is provided by scaled models, in which the drive decreases with increasing system size so that diffusion dominates on local scales, while drift becomes relevant only over distances comparable to the system size. This separation of scales often permits controlled hydrodynamic descriptions and analytical approaches that are not available in the strongly driven case. Since the drive is weak at the microscopic level, the resulting thermodynamic limit can nonetheless differ qualitatively from that of the corresponding model with finite drive, yielding new phases and altered critical properties. When such differences arise, it is natural to ask how the two limits are related, and how one form of collective behavior transforms into the other as the drive is varied.

Several paradigmatic models illustrate this distinction. The weakly asymmetric exclusion process (WASEP)~\cite{Enaud2004}, obtained by scaling the asymmetry of the exclusion process inversely proportional to the system size, provides an interpolation between purely diffusive and strongly driven transport. Its large-deviation functional explicitly connects exact results known for the symmetric and asymmetric exclusion processes, illustrating how weak-driving limits can smoothly bridge equilibrium and nonequilibrium regimes. Another important example is the ABC model~\cite{Evans1998}, where scaling the microscopic bias with system size~\cite{Clincy2003} yields a phase diagram that includes disordered, critical, and phase-separated regimes.  Notably, the critical phase features anomalous long-range correlations~\cite{Gerschenfeld2012} and fluctuations dominated by the lowest Fourier mode~\cite{Gerschenfeld2011}, properties that are not known to appear in the analogous model with finite drive. In a similar spirit, scaled drive and noise formulations have also been utilized to understand the fluctuating hydrodynamics of several systems including active lattice gases~\cite{agranov2021,jose2023} and evolutionary models~\cite{Bhutia2026}. More generally, these cases show that different choices of drive scaling can fundamentally reshape the thermodynamic limit, producing qualitatively new phases, fluctuation patterns, and correlation structures, and providing concrete realizations of critical states that arise only in the weakly driven regime.

%%%%%%%%%%%%%%%%%%%%%%%%%%%%%%%%%%%%%%%%%%%%%%%%%%%%%%%%%%%%%%%%%%%%%
% Ensure \usepackage{array} is in your preamble
\begin{table}[t]
\centering
\renewcommand{\arraystretch}{1.25}
\setlength{\tabcolsep}{5pt}
\begin{tabular}{>{\raggedright\arraybackslash}p{2.6cm} >{\raggedright\arraybackslash}p{4.25cm} >{\raggedright\arraybackslash}p{4.25cm}}
\hline
\textbf{} & \textbf{Unscaled LH model} & \textbf{Scaled LH model} \\
\hline

\textbf{Criticality: correlation}
&
$C(r)\sim m_c^2-b(r/L)^\alpha$; $O(1)$~\cite{Das2000} amplitude with a cusp singularity~; Eq~\eqref{eq:fdpo}.
&
$C(r)\sim L^{-1/2}\cos(2\pi r/L)$~; Eq~\eqref{eq:scaled_critical_correlations}.

\\[1.5mm]

\textbf{Criticality: Fourier spectrum}
&
FDPO with broad multi-mode spectrum $S(\kappa) \sim \kappa^{-(1+\alpha)}$
and $\langle Q_{n}\rangle \sim L^{-\phi}Y(n/L)$ (Eqs~\eqref{eq:density_fm},~\eqref{eq:ulh_fm})~\cite{Kapri2016}; nonlinear hydrodynamics.
&
Single-mode $k_1=2\pi$ dominates ($\langle Q_{1}\rangle\sim L^{-1/4}$,~ higher modes $\sim L^{-1/2}$), effective Fokker-Planck description~(Eq~\eqref{eq:fp}).

\\[1.5mm]

\textbf{Ordered phase: Porod's law}
&
Strong phase separation of particles; strong agreement with Porod's law in particle correlation.
&
Long-range order; Porod's law is recovered for separations larger than the width of the interface.

\\[1.5mm]

\textbf{Disordered phase}
&
Local FDPO correlations within $r<\xi$, with $\xi\sim\Delta^{-\nu}$ ($\nu\simeq0.8$); Eq~\eqref{eq:xi_vs_delta}.
&
Purely exponential decay with $\xi\sim\Delta^{-1/2}$~\cite{Prakash2025}; agreement with fluctuating hydrodynamics 

\\
\hline
\end{tabular}
\caption{ Summary of our main results, contrasting multi-mode fluctuation-dominated phase ordering (FDPO) in the unscaled LH model with single-mode criticality in the scaled model.}
\label{tab:summary}
\end{table}
%%%%%%%%%%%%%%%%%%%%%%%%%%%%%%%%%%%%%%%%%%%%%%%%%%%%%%%%%%%%%%%%%%%%%

The Light-Heavy (LH) model~\cite{Lahiri1997,Lahiri2000,Das2000,Das2001,Chakraborty2016,Chakraborty2017,Chakraborty2019,Mahapatra2020,Khamrai2024,Prakash2025} offers a particularly compelling framework in this context. It describes two particle species coupled to a fluctuating surface and exhibits a rich phase diagram that emerges from the interplay between particle dynamics and surface fluctuations. In its standard formulation, where the external drive and diffusion remain of comparable strength as the system size grows, the model displays fluctuation-dominated phase ordering (FDPO)~\cite{Barma2024}. This is an unusual critical state in which long-range order is present but accompanied by exceptionally strong fluctuations. First discovered in systems of particles sliding on fluctuating surfaces~\cite{Das2000,Das2001}, FDPO has subsequently been identified in a broad range of equilibrium~\cite{Barma2019} and nonequilibrium systems~\cite{Mishra2006,Narayan2007,Dey2012,Das2016,Katyal2020}. Hallmarks of FDPO include a broad distribution of the order parameter that remains finite in the thermodynamic limit, and a cusp singularity in the scaled two-point correlation function, signaling the coexistence of large-scale order and strong fluctuations.

More recent studies~\cite{Prakash2025} have demonstrated that the behavior of the LH model changes qualitatively when the drive is scaled with system size. In this scaled version, the FDPO phase vanishes and is replaced by a distinct critical regime. Although the presence and location of this new phase are known, its nature and its relation to the FDPO observed in the unscaled model remain poorly understood. This leads to several natural questions: What type of critical state appears in the scaled limit? In which respects does it differ from FDPO, and are there any ways in which it is similar to FDPO? 

In this work, we address these questions through a combination of analytical calculations and numerical simulations of the LH model. We first examine the unscaled regime and study the persistence of FDPO-like behavior within the homogeneous phase. In particular, we show that FDPO-like behavior persists on length scales below the correlation length, leading to localized critical behavior and a scaling collapse governed by the correlation length rather than the system size. We next turn to the scaled limit, where we show that the critical state is governed by the first Fourier mode and falls within the class of single-mode-dominated critical states previously identified in the scaled ABC model. Using the theoretical framework developed for the sABC model, we then derive the corresponding scaling form of the two-point static correlation function. A summary of our main results is provided in Table~\ref{tab:summary}.

%%%%%%%%%%%%%%%%%%%%%%%%%%%%%%%%%%%%%%%%%%%%%%%%%%%%%%%%%%%%%%%%%%%%%%%%%%%%%%%%%%%%%%
%%%%%%%%%%%%%%%%%%%%%%%%%%%%%%%%%%%%%%%%%%%%%%%%%%%%%%%%%%%%%%%%%%%%%%%%%%%%%%%%%%%%%%
\section{The model}
\label{sec:model}
%%%%%%%%%%%%%%%%%%%%%%%%%%%%%%%%%%%%%%%%%%%%%%%%%%%%%%%%%%%%%%%%%%%%%%%%%%%%%%%%%%%%%
%%%%%%%%%%%%%%%%%%%%%%%%%%%%%%%%%%%%%%%%%%%%%%%%%%%%%%%%%%%%%%%%%%%%%%%5
\begin{figure}[t!]
\centering
\includegraphics[width=1.0\linewidth] {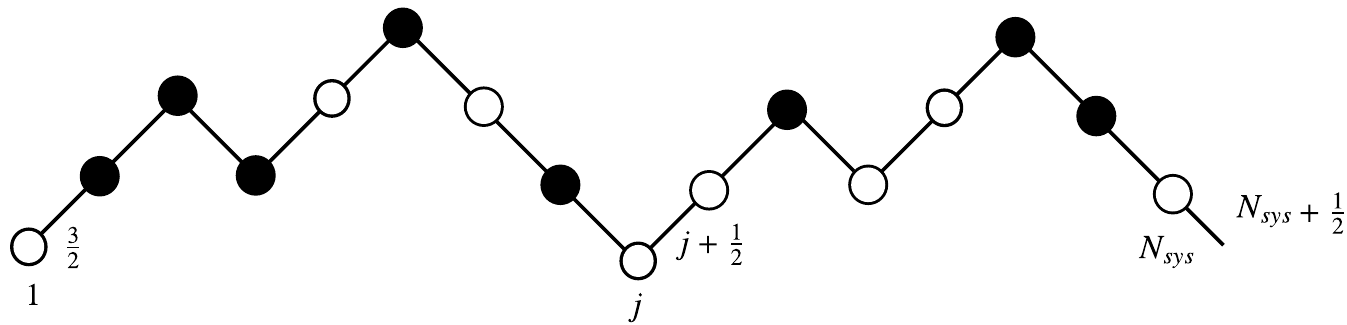}
\caption{Schematic illustration of a Light-Heavy model configuration. Light particles ($\circ$) and heavy particles ($\bullet$) occupy integer lattice sites $j$, whereas up tilts ($/$) and down tilts ($\setminus$) reside on half-integer sites $j + \frac{1}{2}$.}
\label{fig:schematic}
\end{figure}
%%%%%%%%%%%%%%%%%%%%%%%%%%%%%%%%%%%%%%%%%%%%%%%%%%%%%%%%%%%%%%%%%%%%%%%

The LH model, illustrated in Figure~\ref{fig:schematic}, describes a lattice of particles moving on a fluctuating surface, with the dynamics of the particles and the surface mutually influencing one another. Its conceptual origin dates back to the 1970s, when Crowley~\cite{Crowley1971}, in a sedimentation experiment, observed that steel balls settling in turpentine spontaneously formed clusters, even though they had initially been placed at uniform separations. The theoretical description proposed at that time omitted key ingredients such as Brownian motion, elastic interactions, and nonlinearities. Motivated by this, Lahiri and Ramaswamy introduced what is now known as the LH model~\cite{Lahiri1997} in the late 1990s, mapping the experimental system (characterized by two coupled degrees of freedom, particle density and an orientation variable) onto two interacting Ising variables, representing particles and tilt. Besides successfully capturing the clustering observed in the original experiment, the LH model has since emerged as a minimal framework for studying systems with mutually coupled dynamics.

A natural biological setting for coupled dynamics arises in the behavior of proteins and lipids diffused on cell membranes, where the clustering of particles is tied to membrane fluctuations driven by the actin cytoskeleton~\cite{Goswami2008}. Initial theoretical descriptions~\cite{Das2016,Singha2023} treated the membrane as evolving independently, with proteins and lipids merely adapting to the resulting membrane landscape. Ref~\cite{Das2016} predicts macroscopic clusters that are in constant flux, a hallmark of fluctuating dominated phase ordering (FDPO). However, experiments~\cite{Yu2011} demonstrate that membrane-associated proteins can themselves reorganize the cytoskeleton. This suggests that the two species must be modeled as dynamically coupled, as done in the LH model.

%%%%%%%%%%%%%%%%%%%%%%%%%%%%%%%%%%%%%%%%%%%%%%%%%%%%%%%%%%%%%%%%%%%%%%%5
\begin{figure}[t!]
\centering
\includegraphics[width=1.0\linewidth] {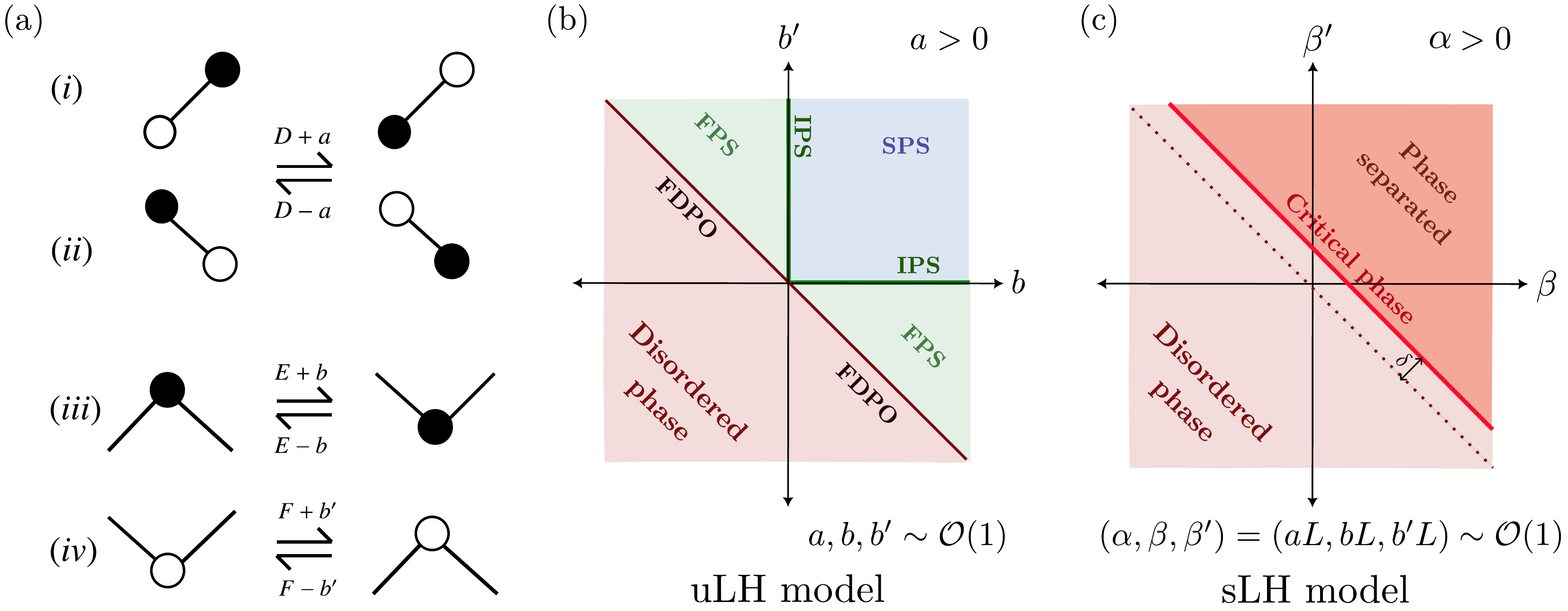}
\caption{(a) A schematic depicting the local update rules of the Light-Heavy (LH) model. (b) An illustration of a two-dimensional cross-section of the phase diagram of the uLH model. This model displays three ordered phases: strong phase separation (SPS); infinitesimal current with phase separation (IPS); and finite current with phase separation (FPS). All three exhibit strong phase separation of particles (i.e., the interfacial region is a few sites wide), along with varying degrees of tilt separation, which is strongest in SPS and weakest in FPS. Beside these, there exist a disordered phase and a distinct regime, Fluctuation Dominated Phase Ordering (FDPO) $(b^{\prime}=-b)$, which separates the ordered and disordered regions. (c) The corresponding phase diagram of the sLH model in two dimensions, which has a comparatively simpler phase structure with three regimes: one with conventional phase separation of both particles and tilts (i.e., the interfacial region has a finite width), a disordered phase, and a critical phase that separates the disordered and phase-separated regions. The locus of this critical phase is shifted from the $\beta^{\prime}=-\beta$ line by an amount $\delta = 2\pi^{2} D^{2}/(\alpha \rho_{0}(1-\rho_{0}))$.}
\label{fig:pd}
\end{figure}
%%%%%%%%%%%%%%%%%%%%%%%%%%%%%%%%%%%%%%%%%%%%%%%%%%%%%%%%%%%%%%%%%%%%%%%
As noted in the previous paragraph, the LH model consists of two interacting Ising-type variables. One set corresponds to particles, which can be either heavy ($\bullet$) or light ($\circ$). The other set represents the local tilt of the surface, which can point up (/ – oriented at an angle $\pi/4$) or down ($\backslash$ – oriented at an angle $3\pi/4$). We encode these degrees of freedom using spin variables $\sigma$ and $\tau$, following the convention specified below:
\\
%%%%%%%%%%%%%%%%%%%%%%%%%%%%%%%%%%%%
\newline
$\sigma_{j} = 
        \begin{array}{cc}
         \Biggl\{ &
          \begin{matrix}
            -1 & \text{Light}  &\circ\\
            +1 & \text{Heavy}  &\bullet    
          \end{matrix}
        \end{array}\hspace{4cm}
\tau_{j+\frac{1}{2}} = 
        \begin{array}{cc}
         \Biggl\{ &
          \begin{matrix}
            -1  &\text{Up}  &/\\
            +1  &\text{Down} &\setminus    
          \end{matrix}
        \end{array}$\\
%%%%%%%%%%%%%%%%%%%%%%%%%%%%%%%%%%%%%%%%%
\\

The two species can equivalently be described in terms of their coarse-grained density fields. These density variables enter the hydrodynamic equations for the system (see section~\ref{sec:Fluctuating hydrodynamics}). We denote by $\rho(x,t)$ and $m(x,t)$ the fields representing the heavy particles and the down-tilts, respectively. 

It is also useful to note that, in what follows, we will work with the two-point correlation function
\[
C_{\lambda}(x)=\langle \lambda_{0}\lambda_{x}\rangle-\langle \lambda\rangle^{2}.
\]
In the unscaled model, we will refer to the spin–spin correlation function with $\lambda=\sigma,\tau$ and $\langle\lambda\rangle=0$ (sections~\ref{sec:phase diagram in uLH},~\ref{sec:scaling}), following the convention used in earlier FDPO studies. In contrast, for the scaled model derived from fluctuating hydrodynamics (section~\ref{sec:Critical behavior}), we will use density fields with $\lambda=\rho,m$, where $\langle \lambda\rangle\neq0$. The two-point correlation functions in the spin and density representations are simply related by
\[
C_{\sigma}(x)=4\bigl(C_{\rho}(x)+\rho_{0}^{2}\bigr)-1,
\]
with an analogous relation holding for the tilt fields.

Microscopically, the particle and tilt variables are placed on two interpenetrating sublattices, forming a configuration that can be viewed as particles sitting on a fluctuating surface as illustrated in Figure~\ref{fig:schematic}. The interaction between the two species is such that the configuration of one species at a given time step determines how the other evolves in the next time step. The detailed dynamical update rules are shown in panel $a$ of Figure~\ref{fig:pd}.

The parameters \(D, E, F\) in the update rules determine the diffusion rates, whereas \(a, b, b'\) represent the bias rates. We note that if the drift is completely turned off by setting \(a = b = b' = 0\), the dynamics decompose into two independent Symmetric Simple Exclusion Processes (SSEPs). Hence, the drift term is precisely what couples the two subsystems, and the ratio of its magnitude to the diffusion rates dictates the structure of the resulting phase diagram. Two versions of the model can be considered: the unscaled LH model and the scaled LH model.

\subsection{Unscaled LH model}
The unscaled LH (uLH) model corresponds to the regime in which the drive and diffusion are of comparable strength, i.e., \(\mathcal{O}(D)\sim\mathcal{O}(a, b, b^{\prime})\). In this limit, the system is far from equilibrium and develops correlations even at local scales. The resulting phase diagram (Fig.~\ref{fig:pd} b) features a range of unconventional phases, such as strong phase separation, FDPO, and others. A detailed account of this phase diagram is presented in the next section.

\subsection{Scaled LH model}
By contrast, the scaled LH (sLH) model is defined by rescaling the drift with the system size so that it becomes weak relative to diffusion, i.e., the parameters \(a, b, b'\) scale as \(\mathcal{O}(1/L)\), while the diffusion rates remain \(\mathcal{O}(1)\). In this case, the dynamics are diffusive at local scales, with drift becoming important only at the scale of the entire system. The system thus lies in a near-equilibrium regime. The sLH limit exhibits a different, less intricate, phase structure (Fig.~\ref{fig:pd} c) compared with the uLH model. The phase diagram is analyzed in detail in section~\ref{sec:pd sLH}.

\vspace{0.5cm}

Monte Carlo simulations of this model are carried out by randomly selecting a particle and a tilt at each microscopic time step \((t, t + \Delta t)\), and then updating the state of the species on either side according to the update rules of the system. Over one unit of time, we perform a number of microscopic updates equal to the total number of lattice sites, ensuring that, on average, each particle and each tilt is updated once. We impose periodic boundary conditions, which guarantee a mean slope of zero. Furthermore, each sublattice is chosen to have an even number of sites.

%%%%%%%%%%%%%%%%%%%%%%%%%%%%%%%%%%%%%%%%%%%%%%%%%%%%%%%%%%%%%%%%%%%%%%%%%%%%%%%%%%%%%%
%%%%%%%%%%%%%%%%%%%%%%%%%%%%%%%%%%%%%%%%%%%%%%%%%%%%%%%%%%%%%%%%%%%%%%%%%%%%%%%%%%%%%%
\section{The unscaled regime}
\label{sec:phase diagram in uLH}
%%%%%%%%%%%%%%%%%%%%%%%%%%%%%%%%%%%%%%%%%%%%%%%%%%%%%%%%%%%%%%%%%%%%%%%%%%%%%%%%%%%%%
As discussed in the previous section, the system is described as being in the unscaled regime when the drift term is of the same order of magnitude as the diffusion term. In this situation, the system is driven far from equilibrium and exhibits a complex phase diagram that is not easily accessible by analytical methods beyond mean-field theory. Nevertheless, the exact hydrodynamic equations obtained from the scaled model remain valid at the mean-field level even in the unscaled regime. A linear stability analysis of these mean-field hydrodynamic equations indicates the presence of an order–disorder transition line at $b + b^{\prime} = 0$ in the phase diagram. This analysis is presented in greater detail in the discussion of the sLH model, where it can be carried out exactly.
%%%%%%%%%%%%%%%%%%%%%%%%%%%%%%%%%%%%%%%%%%%%%%%%%%%%%%%%%%%%%%%%%%%%%%%5
\subsection{Phase diagram in the unscaled regime}
\label{sec:pd}
The phase diagram of the uLH model~\cite{Lahiri2000,Chakraborty2016,Chakraborty2017,Chakraborty2019}, shown in panel $b$ of Figure~\ref{fig:pd}, contains five distinct phases: three ordered phases, one disordered phase, and FDPO. As discussed in Section~\ref{sec:intro}, FDPO exhibits persistent order despite strong fluctuations, leading to dynamic particle clusters whose size scales with the system size.  
Turning now to the three ordered phases: in the LH model, as in any system obeying Kawasaki dynamics, order is achieved via phase separation between species. The degree of this phase separation distinguishes the individual ordered phases. All three are characterized by a strong separation of particles, with domain walls typically only a few lattice sites wide, while the surface ordering varies between them. When the tilts are also strongly separated $(a > 0,\quad 
\{b,b' > 0\})$, the system is in the Strong Phase Separation (SPS) phase. If the tilt degrees of freedom are only very weakly disordered $(a > 0,\quad \{(b,b') : bb'=0,\; b+b' > 0\})$, generating an infinitesimal tilt current, we obtain the Infinitesimal current Phase Separation (IPS) phase. Conversely, when there is a significant amount of surface disorder producing a finite tilt current $(a > 0,\quad \{(b,b') : bb'<0,\; b+b' > 0\})$, the system is in the Finite current Phase Separation (FPS) phase. 

In the disordered phase $(a > 0,\quad 
\{b+b' < 0\})$, the system fails to develop any macroscopic clusters: neither particles nor tilts aggregate into large domains. Instead, it resembles a nearly flat surface with a spatially homogeneous mixture of both particle types. Finally, if we choose $a < 0$, the phase diagram is simply reflected about the $b+b^{\prime}$ axis: SPS now appears at $(a < 0,\quad 
\{b,b^{\prime} < 0\})$, IPS at $(a < 0,\quad \{(b,b^{\prime}) : bb^{\prime}=0,\; b+b^{\prime} < 0\})$, FPS at $(a < 0,\quad \{(b,b^{\prime}) : bb^{\prime}<0,\; b+b' < 0\})$, and the disordered phase at $(a < 0,\quad 
\{b+b^{\prime} > 0\})$, while the FDPO phase remains fixed along $(b+b^{\prime}=0)$.

\subsection{Particle–Tilt Duality}
\label{sec:duality}
The dynamical rules of the LH model, shown in panel $a$ of Figure~\ref{fig:pd}, reveal a symmetry between how particles respond to tilts and how tilts respond to particles. To make this connection clear, we temporarily ignore the fact that tilt dynamics involves two bias rates, $b$ and $b^{\prime}$, encoding a species-dependent response, whereas particle dynamics depends only on a single bias rate $a$. This apparent symmetry between the two types of degrees of freedom hints at an underlying duality, which becomes explicit through the following mapping: replace $/$ with $\bullet$ and $\setminus$ with $\circ$. Under this transformation, the rule $/\bullet\setminus \rightleftharpoons \setminus\bullet/$ is mapped to $\bullet/\circ \rightleftharpoons \circ/\bullet$, and $\setminus\circ/ \rightleftharpoons /\circ\setminus$ is mapped to $\circ\setminus\bullet \rightleftharpoons \bullet\setminus\circ$. From this correspondence, we infer a duality provided that the bias $a$ is mapped to $-b=-b^{\prime}$. Equivalently, any dynamical behavior of the particles at $a=c_{0}$ with fixed $b=b^{\prime}=c_{1}$ is exactly reproduced by the tilts when $b=b^{\prime}=-c_{0}$ and $a=-c_{1}$. Note, however, that this duality holds only when the tilt bias rates $b$ and $b^{\prime}$ are taken to be equal, implying that any mirrored behavior between particles and tilts is restricted to the $a$ versus $b=b^{\prime}$ plane.

\subsection{Implications of duality in the phase diagram}
\label{sec:implications on pd}
%%%%%%%%%%%%%%%%%%%%%%%%%%%%%%%%%%%%%%%%%%%%%%%%%%%%%%%%%%%%%%%%%%%%%%%5
\begin{figure}[t!]
\centering
\includegraphics[width=1.0\linewidth] {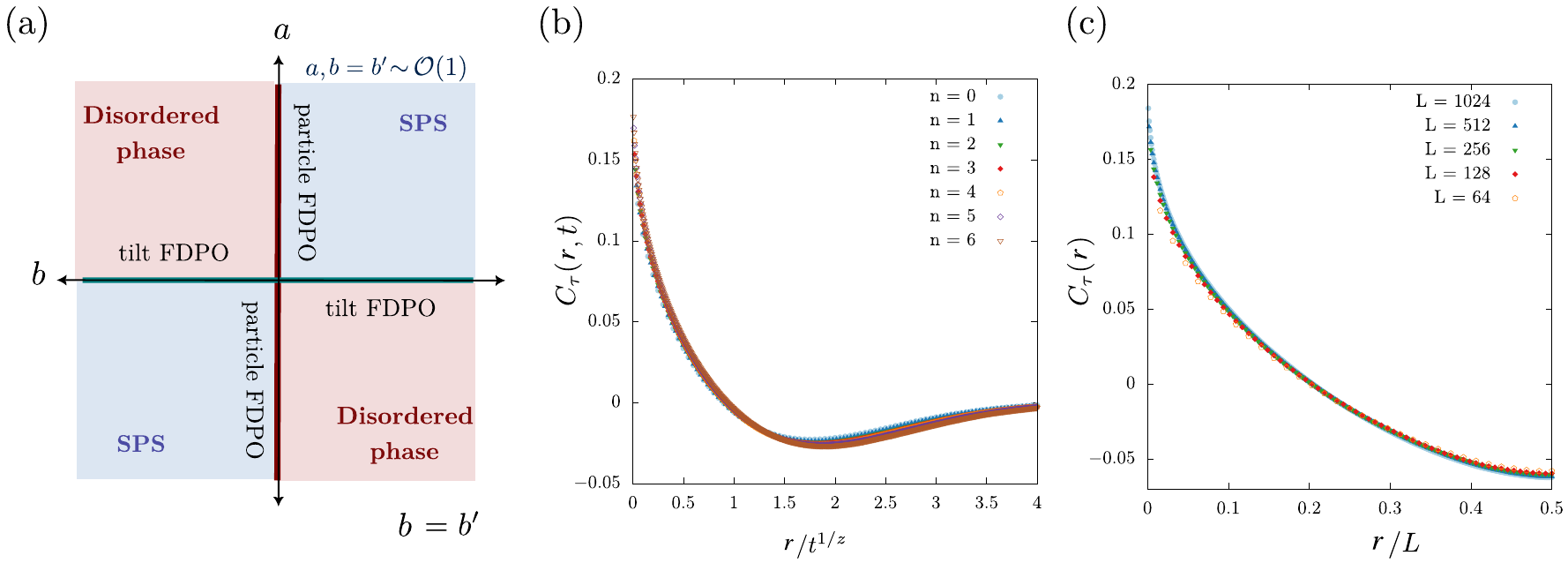}
\caption{(a) A two-dimensional slice of the phase diagram of the uLH model taken along the $b = b^{\prime}$ plane, illustrating FDPO arising from tilts in contrast to that originating from particles. (b) Spin-spin correlation function of the tilts, $C_{\tau}(r, t)$ evaluated at times $t = 400 \times 2^n$ (with $n = 0, \ldots, 6$), demonstrating data collapse under rescaling by $\mathcal{L}(t)=t^{1/z}$ with $z = 2$ (the Edwards–Wilkinson dynamic exponent). (c) Steady-state tilt correlation function $C_{\tau}(r)$ exhibiting collapse when distances are scaled by the system size for $L = 64, 128, 256, 512,$ and $1024$. Simulation parameters: $a = 0$, $b = b^{\prime} = 0.5$.} 
\label{fig:fdpo_tilt}
\end{figure}
%%%%%%%%%%%%%%%%%%%%%%%%%%%%%%%%%%%%%%%%%%%%%%%%%%%%%%%%%%%%%%%%%%%%%%%
An immediate implication of this duality is the possibility of an FDPO of tilts within the system in a suitable region of the phase diagram, serving as the dual counterpart to the FDPO of particles. In the $a$ versus $b=b^{\prime}$ plane, FDPO of particles occurs along the $a\neq 0$ and $b=b^{\prime}=0$ axis. These parameter values correspond to Edwards–Wilkinson (EW) surface dynamics, and the FDPO phase that appears there is known to be characterized by a correlation function with a cusp exponent of $0.5$. By duality, this implies a corresponding FDPO of tilts, also with a cusp exponent of $0.5$, along the $a=0$ and $b=b^{\prime}\neq 0$ axis. This is precisely what we find. Figure~\ref{fig:fdpo_tilt} presents Monte Carlo data for the tilt–tilt correlation function. In panel (b), which corresponds to FDPO during the coarsening regime, we find the anticipated cusp exponent $\alpha=1/z=0.5$, while panel (c) confirms the appropriate scaling with system size characteristic of the steady state.

FDPO of the particles is known to demarcate a strongly phase-separated (SPS) state of particles and tilts, occurring at $(a>0,\; b=b^{\prime}>0)$ and $(a<0,\; b=b^{\prime}<0)$, from a disordered regime at $(a>0,\; b=b^{\prime}<0)$ and $(a<0,\; b=b^{\prime}>0)$, respectively, in the $a$–$b=b^{\prime}$ plane. In a completely analogous way, the FDPO of the tilts separates SPS at $(b=b^{\prime}>0,\; a>0)$ and $(b=b^{\prime}<0,\; a<0)$ from the disordered phase at $(b=b^{\prime}>0,\; a<0)$ and $(b=b^{\prime}<0,\; a>0)$, again along the same plane. Since there is no parameter $a^{\prime}$ appearing in the dynamics, the FDPO of particles arising from Kardar–Parisi–Zhang (KPZ)-type surface dynamics does not admit such a dual description.

Although the appearance of FDPO of tilts is not surprising since it follows directly from the symmetries of the dynamics, it is still useful, as it offers insight into the organization of the phase diagram. To conclude, we stress that all of our preceding discussion of the phase diagram strictly concerns the situation in which all bias parameters $a, b, b^{\prime}$ are kept unscaled relative to the diffusion rate.

%%%%%%%%%%%%%%%%%%%%%%%%%%%%%%%%%%%%%%%%%%%%%%%%%%%%%%%%%%%%%%%%%%%%%%%%%%%%%%%%%%%%%
%%%%%%%%%%%%%%%%%%%%%%%%%%%%%%%%%%%%%%%%%%%%%%%%%%%%%%%%%%%%%%%%%%%%%%%%%%%%%%%%%%%%%
\section{FDPO-like behavior in the disordered phase}
\label{sec:scaling}
%%%%%%%%%%%%%%%%%%%%%%%%%%%%%%%%%%%%%%%%%%%%%%%%%%%%%%%%%%%%%%%%%%%%%%%%%%%%%%%%%%%%%
%%%%%%%%%%%%%%%%%%%%%%%%%%%%%%%%%%%%%%%%%%%%%%%%%%%%%%%%%%%%%%%%%%%%%%%5
\begin{figure}[t!]
\centering
\includegraphics[width=1.0\linewidth] {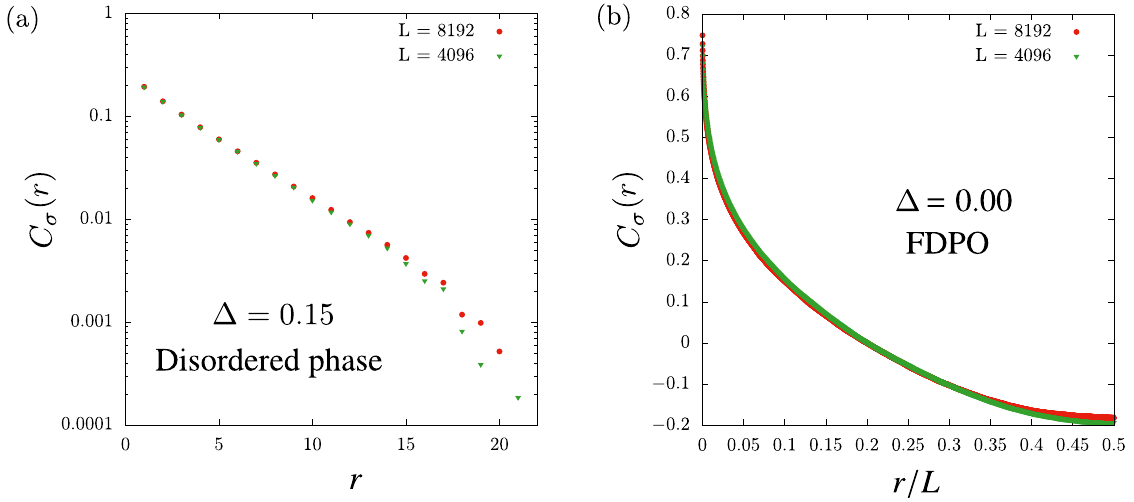}
\caption{(a) The particle–particle correlation function in the deeply disordered regime as a function of separation for system sizes $4096$ and $8192$. Characteristic of disordered behavior, the results exhibit no dependence on system size, and the two data sets lie on top of each other. (b) The corresponding data at a point displaying FDPO plotted versus scaled separation. The collapse of the data demonstrates the system-size scaling associated with FDPO.}
\label{fig:cor_unscaled}
\end{figure}
%%%%%%%%%%%%%%%%%%%%%%%%%%%%%%%%%%%%%%%%%%%%%%%%%%%%%%%%%%%%%%%%%%%%%%%
In the previous section, we discussed FDPO in both particles and tilts, which manifests along the critical locus of the uLH model. Here, we move away from this critical locus and instead focus on the homogeneous regime of the unscaled model. We will show that within this domain, the system exhibits FDPO-like behavior on local length scales, closely mirroring systems displaying conventional critical properties. 

Fluctuation-Dominated Phase Ordering, introduced in Section~\ref{sec:intro}, shares a number of features with conventional critical phenomena, while also exhibiting important distinctions. These similarities and differences are reflected in the behavior of the equal-time two-point correlation function,
\begin{equation}
    C(r)=m_{c}^{2}-b_{ss}\biggl(\frac{r}{L}\biggr)^{\alpha}
    \label{eq:fdpo}
\end{equation}
which serves as a useful probe of the ordering characteristics. The correlation function exhibits several traits familiar from standard critical behavior, such as a power-law dependence and singularities, but also departs from it in important respects. The most significant difference is that FDPO is associated with system-size-dependent scaling reminiscent of standard phase-ordering kinetics, whereas in ordinary criticality one expects scale-free fluctuations that are independent of system size in the thermodynamic limit. The resemblances between FDPO and ordinary critical systems naturally prompt the question of how far this analogy can be pursued.

A well-known characteristic of systems with standard critical behavior is that hallmarks of criticality persist even when the system is not exactly at the critical point. In particular, within the homogeneous phase, where correlations are short-ranged and are typically captured by an exponentially decaying correlation function, certain notable features arise. If we consider a subsystem in this phase whose linear size is smaller than the correlation length, then any two particles within this subsystem effectively "fail to notice" that the overall system is off criticality. As a result, their behavior is as though the system were still at the critical point. This manifests itself in the two-point correlation function as well, such that

\begin{equation}
C(r) \sim
\frac{1}{r^{d-2+\eta}}  \qquad r<\xi.
\label{eq:c1}
\end{equation}

We seek to determine whether this phenomenon (well known from systems with conventional critical behavior) also arises in their FDPO counterparts, using the uLH model as our case study. Put differently, do subsystems whose linear size is smaller than the correlation length in the homogeneous phase of the uLH model exhibit behavior analogous to that characteristic of FDPO? To explore this issue, we will calculate two-point correlation functions in the homogeneous phase of the unscaled LH model using Monte Carlo simulations and investigate whether
\begin{equation}
C(r) \sim
m_{c}^{2}-b_{ss}\bigg(\frac{r}{\xi}\bigg)^{\alpha}; \qquad r<\xi
\label{eq:fdpocases}
\end{equation}
with $\xi$ replacing $L$ in Eq.~\eqref{eq:fdpo} holds away from criticality.
%%%%%%%%%%%%%%%%%%%%%%%%%%%%%%%%%%%%%%%%%%%%%%%%%%%%%%%%%%%%%%%%%%%%%%%5
\begin{figure}[t!]
\centering
\includegraphics[width=1.0\linewidth] {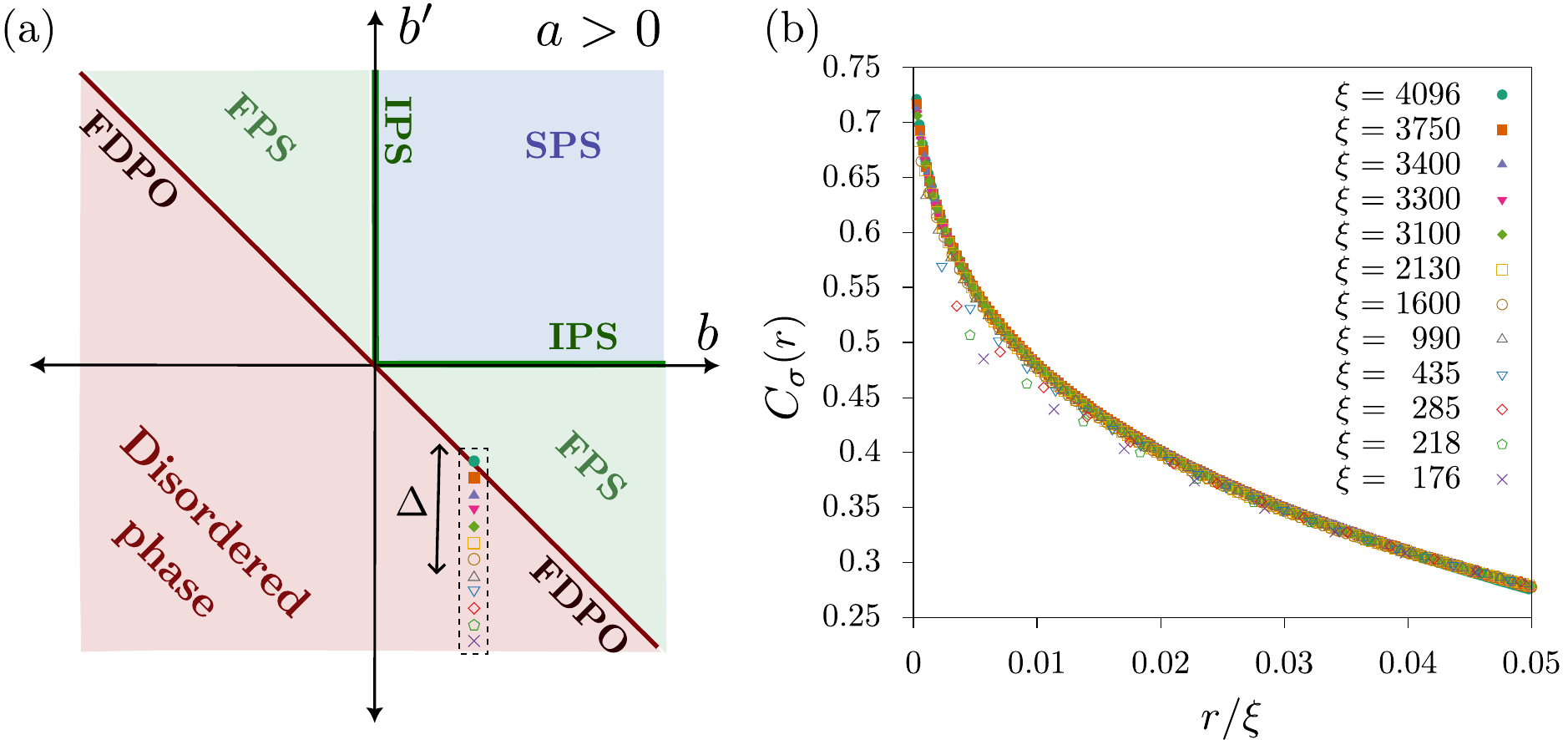}
\caption{(a) Phase diagram of the uLH model with markers indicating the locations in the homogeneous region of phase space that were sampled for data. (b) Data collapse of the corresponding spin–spin correlation functions after rescaling by the length scale $\xi$ for a system of size $L = 4096$. The parameters are $a=0.4$, $b=0.2$, and $b'=-0.2$, with $\Delta = 0,\ 4\times10^{-5},\ 6\times10^{-5},\ 8\times10^{-5},\ 10^{-4},\ 3\times10^{-4},\ 5\times10^{-4},\ 10^{-3},\ 3\times10^{-3},\ 5\times10^{-3},\ 7\times10^{-3},\ 9\times10^{-3}$.} 
\label{fig:discor_scaled}
\end{figure}
%%%%%%%%%%%%%%%%%%%%%%%%%%%%%%%%%%%%%%%%%%%%%%%%%%%%%%%%%%%%%%%%%%%%%%%
%%%%%%%%%%%%%%%%%%%%%%%%%%%%%%%%%%%%%%%%%%%%%%%%%%%%%%%%%%%%%%%%%%%%%%%5
\begin{figure}[t!]
\centering
\includegraphics[width=1.0\linewidth] {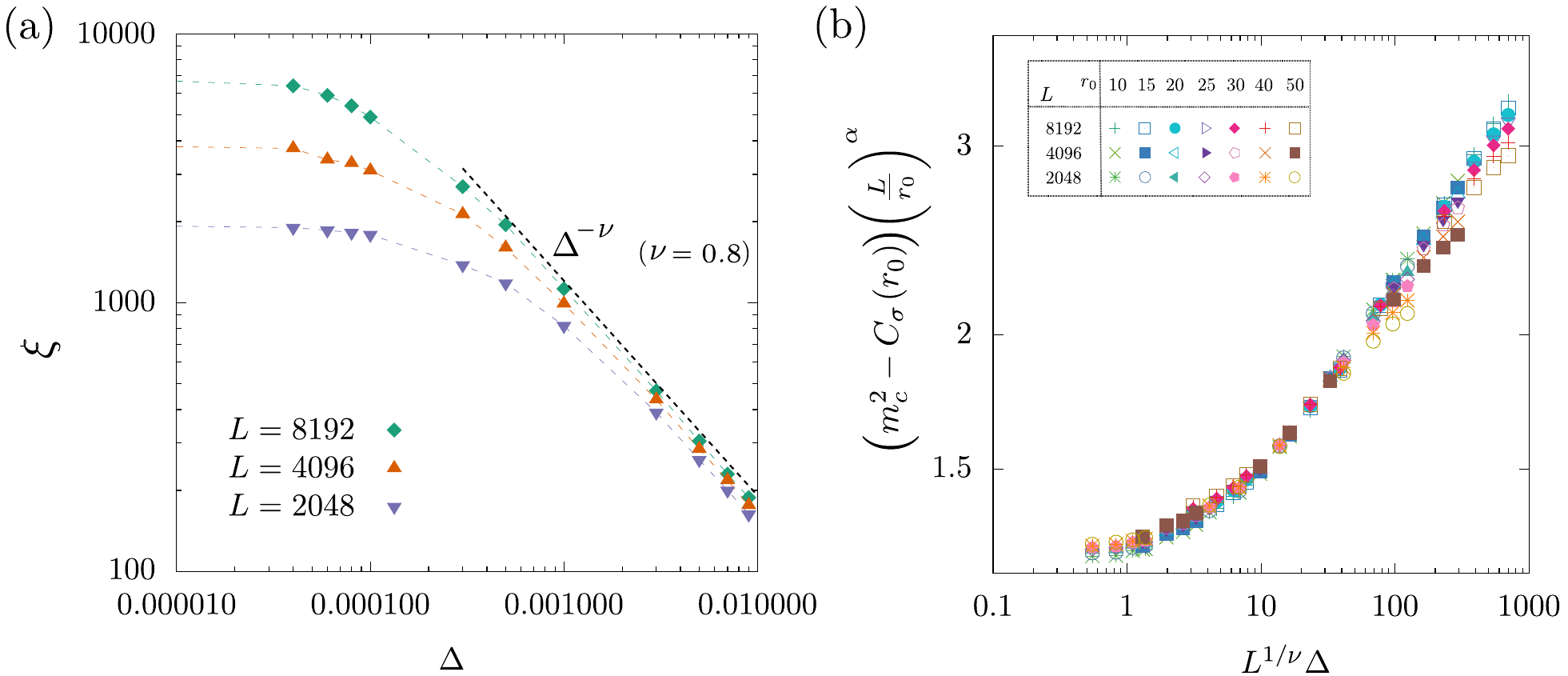}
\caption{(a) The correlation length $\xi$, obtained from the spin–spin correlation functions displayed in Fig.~\ref{fig:discor_scaled} for $L=4096$ (together with analogous data for $L=8192$ and $L=2048$), is shown as a function of the distance $\Delta$ from FDPO. Away from FDPO, $\xi$ exhibits power-law behavior, $\xi \sim \Delta^{-\nu}$ with an exponent $\nu = 0.8$. Consequently, as the system is tuned closer to FDPO $(\Delta \to 0)$, the correlation length increases algebraically until it becomes of the order of the system size, at which point its further growth is suppressed by finite-size effects and it saturates at $\xi \sim L$. (b) The corresponding correlation functions are rescaled according to Eq.~\eqref{eq:scaling function}, producing a data collapse consistent with the expected scaling form. Note: the values of $r_{0}$ used to assess the scaling form are chosen such that $r_{0}\to\infty$, $L\to\infty$, but $r_{0}/L\to 0$.} 
\label{fig:Lfdpdo_kpz}
\end{figure}
%%%%%%%%%%%%%%%%%%%%%%%%%%%%%%%%%%%%%%%%%%%%%%%%%%%%%%%%%%%%%%%%%%%%%%%
We begin by comparing the behavior of the system at FDPO with a point sufficiently far into the homogeneous regime (Figure~\ref{fig:cor_unscaled}). Panel $a$ shows that, well into the disordered regime, the correlation function decays in a purely exponential manner with a correlation length that does not depend on the system size. By comparison, panel $b$ illustrates the FDPO regime, where the correlations clearly scale with the system size and display a distinct cusp singularity as $r/L \rightarrow 0$. Our goal is to show that, upon moving away from FDPO into the disordered phase, the system nevertheless exhibits FDPO-like behavior within each length scale $\xi$. To this end, we adjust the parameters to lie sufficiently close to the critical line so that the correlation length becomes large, while still maintaining $\xi \ll L$, allowing the system to display potentially interesting, critical-like features.

The most straightforward way to extract the correlation length $\xi$ is to rescale the off-critical data so that it collapses onto the master curve obtained at FDPO from systems of different sizes (see Figure~\ref{fig:cor_unscaled}), taking the required scaling factors as the respective values of $\xi$. This strategy, however, works reliably only close to the critical line where $\xi \sim L$. Deeper in the off-critical regime ($\xi \ll L$), the intercept $m_c^2$ develops an explicit dependence on $\xi$. This is analogous to the finite-size behavior at FDPO, where $m_c^2$ is $L$-dependent, so the data collapse becomes less accurate near the origin for smaller systems.

A second strategy is to perform direct fits using Eq.~\eqref{eq:fdpocases}. This can serve either as a consistency check on the $\xi$ values inferred from the data collapse or as an alternative procedure in which $\xi$ is fitted simultaneously with $m_c^2$ and $b_{ss}$. The core difficulty is that the prefactor $b_{ss}$ can absorb any unknown proportionality constants. As a result, the fit can accommodate several different choices of $\xi$, so the extracted values are never uniquely fixed. On top of this, Eq.~\eqref{eq:fdpocases} is expected to be asymptotically valid only in the limit $r/\xi \to 0$ with $r \to \infty$ and $\xi \to \infty$. For larger $r/\xi$, one anticipates corrections to Eq.~\eqref{eq:fdpocases}, as expected even for FDPO.

In summary, neither method (whether relying on a scaling collapse that is checked by a fit or treating $\xi$ as an adjustable parameter within the fit) provides a strict, standalone determination of the correlation length. A more robust approach would be to avoid the ambiguities associated with a non-independent $\xi$ altogether and instead work with a parameter intrinsic to the system that does not require explicit determination.
Such a natural parameter in this problem is $\Delta$, defined as the distance in parameter space from the critical locus. We can therefore reformulate the problem as follows. Consider again Eq.~\eqref{eq:fdpocases}. For a given separation $r_{0}$, it effectively states that
\begin{equation}
\frac{m_c^2-C(r_0)}{r_{0}^{\alpha}}
\sim
\begin{cases}
\dfrac{1}{\xi^\alpha}, & \xi \ll L \quad (\Delta \gg \Delta_c),\\[6pt]
\dfrac{1}{L^\alpha}, & \xi \sim L \quad (\Delta \ll \Delta_c),
\end{cases}
\end{equation}
where $\Delta_{c}$ denotes the crossover scale. Assuming 
\begin{equation}
    \xi\sim\Delta^{-\nu},
    \label{eq:xi_vs_delta}
\end{equation}
the crossover point $\Delta_{c}$ follows from the condition $L\sim\xi$ at $\Delta_{c}$. At a distance $\Delta$ from the FDPO regime, we can then propose
\begin{equation}
\begin{aligned}
\frac{m_{c}^{2}-C(r_{0})}{r_{0}^{\alpha}}
&=\frac{1}{L^{\alpha}}f\biggl(\frac{\Delta}{\Delta_{c}}\biggr)\\
&=\frac{1}{L^{\alpha}}f\!\left(\Delta L^{1/\nu}\right),
\label{eq:scaling function}
\end{aligned}
\end{equation}
where $f$ is the scaling function. In this context, $\nu$ can be obtained either by examining how the proposed $\xi$ varies with $\Delta$ in the regime $\Delta\gg\Delta_{c}$ (Figure~\ref{fig:Lfdpdo_kpz}~a) or through a trial-and-error approach to identify the value that produces the best data collapse. We used $\nu = 0.8$, $\alpha = 0.25$, and $m_c^{2} = 0.88$ to produce the data collapse shown in Figure~\ref{fig:Lfdpdo_kpz}. Equation~\eqref{eq:scaling function} is expected to be valid for those $r_{0}$ such that $r/L\to0$, with $r_{0}\to\infty$ and $L\to\infty$. The outcomes of this analysis are shown in Figure~\ref{fig:Lfdpdo_kpz}~b.

This collapse of correlation functions in the homogeneous phase when distances are rescaled by an appropriate characteristic length implies that FDPO phenomenology can also appear locally within the disordered phase on length scales smaller than $\xi$. In this regime, subsystems of size $\xi$ behave as if at the FDPO point: the correlation function has a cusp singularity and obeys the FDPO scaling form, even though the full system is globally disordered and shows exponentially decaying correlations at larger distances. This parallels conventional critical phenomena, where subsystems smaller than the correlation length are insensitive to deviations from criticality. Our findings apply to one particular trajectory approaching FDPO, in which the surface fluctuations follow KPZ dynamics. It remains to analyze the neighborhood of the critical point controlled by EW-type surface fluctuations, as well as other trajectories that approach criticality governed by KPZ-type behavior. The above results, in particular the plot shown in panel (a) of Figure~\ref{fig:Lfdpdo_kpz}, also have implications for the nature of the transition at FDPO. It demonstrates that, as FDPO is approached, the system exhibits a diverging correlation length. Taken together with the observation that the order parameter (namely, the first Fourier mode of the particle density) undergoes a discontinuous jump at the transition from the disordered phase to the ordered phases (with SPS, IPS, and FPS all being strongly phase-separated in terms of particle distribution), this indicates that the disorder–order transition associated with FDPO in the LH model is of mixed order. Taken together with the earlier findings for the Truncated Inverse-Square Distance Ising (TIDSI) model~\cite{Barma2019}, this leads to the suggestion that FDPO may generally correspond to a mixed-order transition.

%%%%%%%%%%%%%%%%%%%%%%%%%%%%%%%%%%%%%%%%%%%%%%%%%%%%%%%%%%%%%%%%%%%%%%%%%%%%%%%%%%%%%%
\section{Beyond the unscaled regime}
\label{sec:phase diagram outside uLH}
%%%%%%%%%%%%%%%%%%%%%%%%%%%%%%%%%%%%%%%%%%%%%%%%%%%%%%%%%%%%%%%%%%%%%%%%%%%%%%%%%%%%%
So far, our analysis has focused exclusively on the unscaled model, in which the drive and diffusion are of the same order. We now turn to regimes beyond this setting. In particular, we consider cases where the bias parameters are reduced relative to the diffusion rate. When the drive rates scale as $\mathcal{O}(1/L)$ in comparison to the diffusion rates, we arrive at the scaled model. In what follows, we summarize the known results for the fluctuating hydrodynamics and phase diagram of this model.

\subsection{Fluctuating hydrodynamics of the scaled model}
\label{sec:Fluctuating hydrodynamics}
%%%%%%%%%%%%%%%%%%%%%%%%%%%%%%%%%%%%%%%%%%%%%%%%%%%%%%%%%%%%%%%%%%%%%%%%%%%%%%%%%%%%%%
Within this scaled setting, since the bias is of order $\mathcal{O}(1/L)$ compared to the diffusion rates, the system appears effectively diffusive when observed on local spatial scales. This near-equilibrium character of the scaled Light–Heavy model permits an exact analytical treatment. This is because diffusion dominates up to certain characteristic length scales in the system, which in turn allows one to factorize correlators up to this scale while coarse-graining. In the scaled limit, this factorization remains exact to first order in inverse system size. The same set of equations derived from this correlator factorization can also be applied to the unscaled model; however, in that case, they are valid only at the level of a mean-field approximation. Exploiting this, we can derive the long-time, large-scale evolution of locally conserved quantities, such as the local particle and tilt densities in the model, which govern the hydrodynamic behavior of the system. Because of these conservation laws, long-wavelength modes relax slowly, and their dynamics are largely insensitive to microscopic details. Thus, starting from the microscopic lattice dynamics, one can reformulate the evolution of the system in terms of coarse-grained density fields that vary smoothly in space and time. This hydrodynamic description is strictly accurate only in the limit of an infinite system size. For finite systems, one must supplement the hydrodynamic equations with a stochastic noise term, leading to a fluctuating hydrodynamic description that faithfully captures the behavior of the system.\\
The fluctuating hydrodynamics of a lattice system can be systematically derived from the microscopic model using methods such as the Lef\`evre–Biroli approach~\cite{Andreanov2006,Lefevre2007}, which is formally equivalent to the Doi–Peliti formalism~\cite{peliti1985,Doi1976,Doi1976b}. In this framework, one begins with the microscopic local update rules and constructs a path-integral representation to obtain the fluctuating hydrodynamic theory. For a system evolving between prescribed initial and final configurations, there exist many possible trajectories in configuration space, each occurring with a different probability. The local update rules fix the weight of each such trajectory. Using these rules, one can express the probability of observing a given history of particle and tilt configurations as a path integral. The associated large-deviation functional coincides with the one obtained via Macroscopic Fluctuation Theory (MFT), starting directly from the fluctuating hydrodynamic equations. This equivalence makes it possible to derive the fluctuating hydrodynamics of the system directly from its microscopic dynamics.\\
The exactness of these equations relies on the existence of local equilibrium in the system. This requirement is met in equilibrium models such as the Symmetric Simple Exclusion Process (SSEP). In driven systems like the Asymmetric Simple Exclusion Process (ASEP), local equilibrium is generally not assured, except possibly at special points in parameter space where some compensating mechanism of configuration currents (such as pairwise balance) holds. When the drive is instead scaled as $\mathcal{O}(1/L)$, as in the Weakly Asymmetric Simple Exclusion Process (WASEP) and the sLH model, the dynamics become diffusive at coarse-grained scales. In this weak-drive regime, the fluctuating hydrodynamic description derived by this method is asymptotically exact, up to corrections of order $\mathcal{O}(1/L)$. The corresponding fluctuating hydrodynamic equations for the sLH model obtained in this way~\cite{Prakash2025} are,
%%%%%%%%%%%%%%%%%%%%%%%%%%%%%%%%%%%%%%%%%
\begin{equation}
 \frac{\partial \rho(x,t)}{\partial t}    = D \frac{\partial^{2} \rho(x,t)}{\partial x^{2}} +\frac{\partial}{\partial x} \bigg[2\alpha\rho(x,t)(1-\rho(x,t))(1-2m(x,t))\bigg] + \frac{\partial \eta_{\rho(x,t)}}{\partial x},
 \label{hyd1}
 \nonumber
\end{equation}
%%%%%%%%%%%%%%%%%%%%%%%%%%%%%%%%%%%%%%%%%%%
\begin{equation}
 \frac{\partial m(x,t)}{\partial t}    = D \frac{\partial^{2} m(x,t)}{\partial x^{2}} +\frac{\partial}{\partial x} \bigg[2m(x,t)\bigl(1-m(x,t)\bigr)\bigl(\bigl(\beta+\beta^{\prime}\bigr)\rho(x,t)-\beta^{\prime}\bigr)\bigg] + \frac{\partial \eta_{m}(x,t)}{\partial x},
 \label{eq:hyd2}
\end{equation}
%%%%%%%%%%%%%%%%%%%%%%%%%%%%%%%%%%%%%%%%%%%%%%
where $\rho(x,t)$ and $m(x,t)$ denote the coarse-grained density fields of the heavy particle and the down-tilt, respectively, as discussed earlier. The noise correlations are given by
%%%%%%%%%%%%%%%%%%%%%%%%%%%%%%%%%%%%%%%%%%%%%
\begin{eqnarray}
 \langle \eta_{\rho}(x,t)  \eta_{\rho}(x^{\prime},t^{\prime})\rangle &=& \frac{2 D \rho(x,t) (1-\rho(x,t))}{L} \delta (x-x^{\prime}) \delta (t-t^{\prime}),
 \nonumber\\
 \langle \eta_{m}(x,t)  \eta_{m}(x^{\prime},t^{\prime})\rangle &=& \frac{2 D m(x,t) (1-m(x,t))}{L} \delta (x-x^{\prime}) \delta (t-t^{\prime}),
 \nonumber\\
 \langle \eta_{\rho}(x,t)  \eta_{m}(x^{\prime},t^{\prime})\rangle &=& 0.
 \label{eq:noise_correlations}
\end{eqnarray}

Here $\alpha, \beta,$ and $\beta^{\prime}$ are the rescaled bias rates, given by $\alpha = aL$, $\beta = bL$, and $\beta^{\prime} = b^{\prime}L$ and we have set $D=E=F$. A linear stability analysis of Eq.~\eqref{eq:hyd2} leads to
\begin{equation}
    \alpha(\beta+\beta^{\prime}) = \frac{k_{n}^{2}D^2}{2\rho_{0} (1-\rho_{0})},
    \label{eq:lsa1}
\end{equation}
where $\rho_{0}$ is the global density is the heavy particles and $k_{n}=2\pi n$ represents the $n$-th Fourier mode.
From Eq.~\eqref{eq:lsa1}, one readily sees that the instability in the scaled model first sets in at the lowest mode, $n=1$, which gives
\begin{equation}
    \alpha(\beta+\beta^{\prime}) = \frac{2\pi^{2}D^2}{\rho_{0} (1-\rho_{0})},
    \label{eq:lsa2}
\end{equation}
as the phase boundary separating the disordered and ordered regimes in the sLH model. The relation in Eq.~\eqref{eq:lsa1} likewise applies at the mean-field level for the uLH model, where it implies that $b+b^{\prime}=0 \quad (L\to\infty)$ marks the transition between the ordered and disordered phases in the unscaled model.
%%%%%%%%%%%%%%%%%%%%%%%%%%%%%%%%%%%%%%%%%%%%%%%%

\subsection{Phase diagram of the scaled LH model}
\label{sec:pd sLH}
In contrast to the unscaled model, which exhibits a variety of exotic phases, the scaled model yields a comparatively simple phase diagram. It contains three distinct phases: an ordered phase, a disordered phase, and a critical phase located along the line specified by Eq.~\eqref{eq:lsa2}. In the sLH model, the ordered phase is characterized by phase separation with finite-width interfaces, which differs from the behavior observed in the uLH model. The second is a disordered phase which, although qualitatively similar to that of the uLH model, is more amenable to analytical treatment due to the near-equilibrium nature of the scaled system. The third is a critical phase that has not yet been fully characterized; it lies between the ordered and disordered phases and constitutes one of the main focuses of this study. This critical phase will be examined in detail in a later section. The corresponding phase diagram for the sLH model is shown in panel $c$ of Figure~\ref{fig:pd}.
%%%%%%%%%%%%%%%%%%%%%%%%%%%%%%%%%%%%%%%%%%%%%%%%%%%%%%%%%%%%%%%%%%%%%%%%%%%%%%%%%%%%%
\section{Critical behavior - from unscaled to scaled}
\label{sec:Critical behavior}
%%%%%%%%%%%%%%%%%%%%%%%%%%%%%%%%%%%%%%%%%%%%%%%%%%%%%%%%%%%%%%%%%%%%%%%%%%%%%%%%%%%%%%
%%%%%%%%%%%%%%%%%%%%%%%%%%%%%%%%%%%%%%%%%%%%%%%%%%%%%%%%%%%%%%%%%%%%%%%

%%%%%%%%%%%%%%%%%%%%%%%%%%%%%%%%%%%%%%%%%%%%%%%%%%%%%%%%%%%%%%%%%%%%%%%

The uLH model, as described in the previous sections, exhibits FDPO along the critical surface, whereas the critical phase in the scaled model has not yet been fully characterized. Linear stability analysis allows us to estimate the location of this surface, but its precise nature is still unclear. When we compute the usual one-point and two-point functions employed to characterize a phase as FDPO, they instead reveal a different type of behavior. Indeed, the replacement of FDPO by a different kind of behavior when system parameters are varied is not unprecedented; it has already been observed in the TIDSI model~\cite{Barma2019,Sadhukhan2025}. The TIDSI model is an Ising system with long-range interactions that act only within domains of like spins. There, it has been demonstrated that the critical behavior crosses over from FDPO to a more conventional power-law–correlated regime as temperature and coupling strength are tuned. In our case, we similarly observe the disappearance of FDPO when the drive is adjusted relative to the diffusion rates.

\begin{figure}[t!]
\centering
\includegraphics[width=1.0\linewidth] {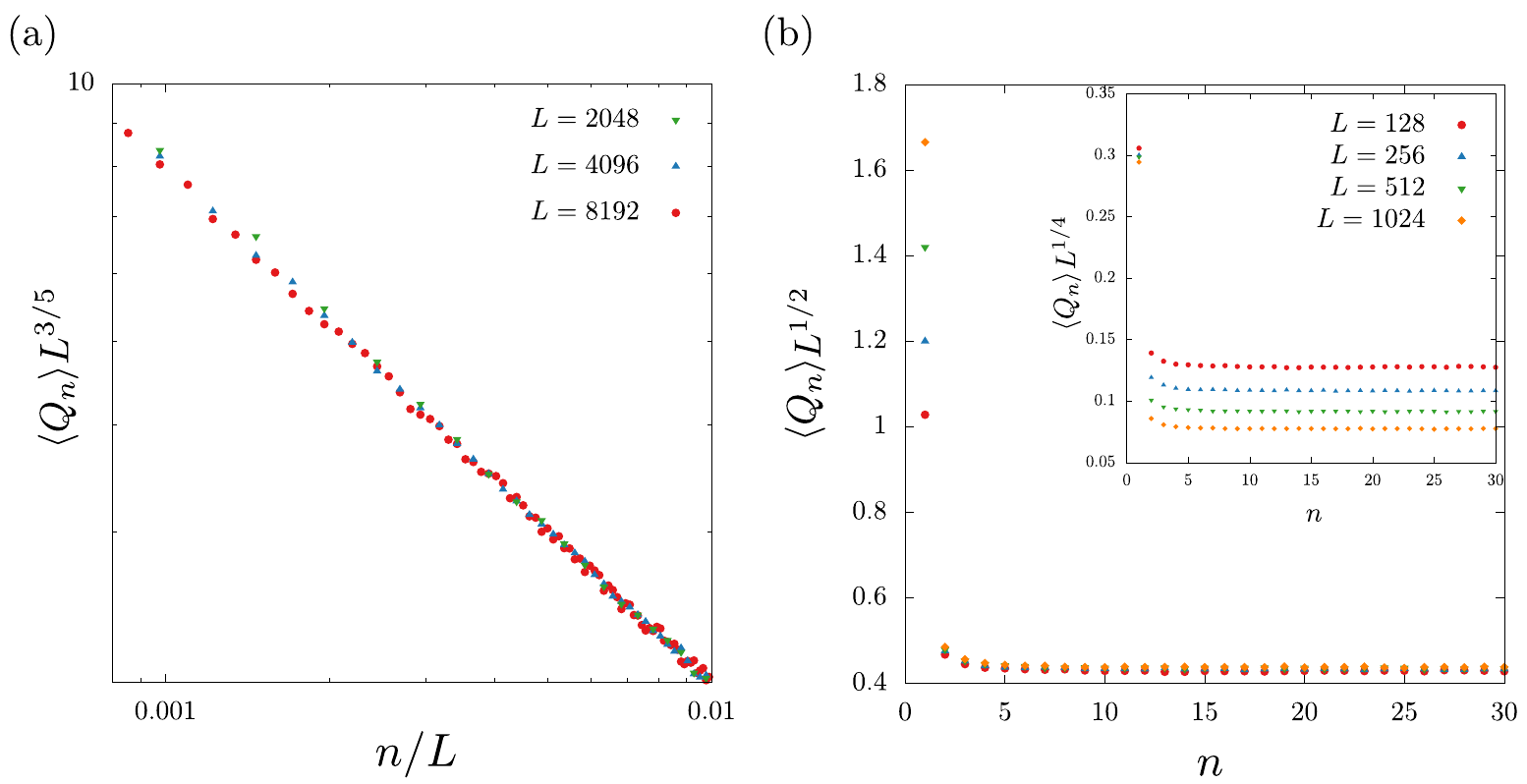}
\caption{(a) Fourier mode of heavy particle density, $Q_n$ as a function of $n$ in the uLH model. The modes are known to follow the scaling form $Q_{nL}\sim L^{-\phi}Y(n/L)$~\cite{Kapri2016}. This figure simply serves as a reminder of this well-established result, so that we can compare it with the behavior in the scaled model. Parameters used are $a=0.4, b=0.2, b^{\prime}=-0.2$. (b) Corresponding plot in the sLH model at $\alpha =\pi/\sqrt{2},~\beta = \beta^{\prime}=\pi \sqrt{2}$. The plot indicates that all modes except $n=1$ decay as $1/\sqrt{L}$ in the thermodynamic limit, whereas the $n=1$ mode (inset) varies slowly, scaling as $1/\sqrt[4]{L}$.} 
\label{fig:crossover}
\end{figure}

To gain a clearer understanding of the critical phase in the scaled model, we analyze the Fourier modes (one-point function) of the system, defined as
\begin{equation}
Q_{n}=\biggl\lvert\frac{1}{L}\sum_{j=1}^{L}\frac{1+\sigma_{j}}{2} e^{i\kappa_{n} j}\biggr\rvert, \qquad \kappa_{n}=\frac{2\pi n}{L}.
    \label{eq:density_fm}
\end{equation}
In the FDPO regime, it has been demonstrated that the dynamics are driven by multiple modes, with a subset of them remaining finite in the thermodynamic limit. In real space, this manifests as dynamically evolving clusters whose size is comparable to the system size, and as a two-point correlation function of $\mathcal{O}(1)$ that scales with system size and exhibits a cusp singularity. It has also been shown that the Fourier modes of the unscaled model display a non-trivial scaling behavior~\cite{Kapri2016},
\begin{equation}
Q_{n}\sim L^{-\phi}Y(n/L),
    \label{eq:ulh_fm}
\end{equation}
where $Y$ denotes the corresponding scaling function and the exponent $\phi$ takes the values $2/3$ (EW) and $3/5$ (KPZ) in the uLH model. We contrast this with the scaled model (Figure~\ref{fig:crossover} b), which features a comparatively simpler mode structure. In this case, no mode remains finite in the thermodynamic limit. The first mode, however, decays more slowly with system size, scaling as $Q_{1}\sim1/\sqrt[4]{L}$, whereas the higher modes decay more rapidly as $Q_{n}\sim1/\sqrt{L} \quad (n\neq 1)$. 

\begin{figure}[t!]
\centering
\includegraphics[width=1.0\linewidth] {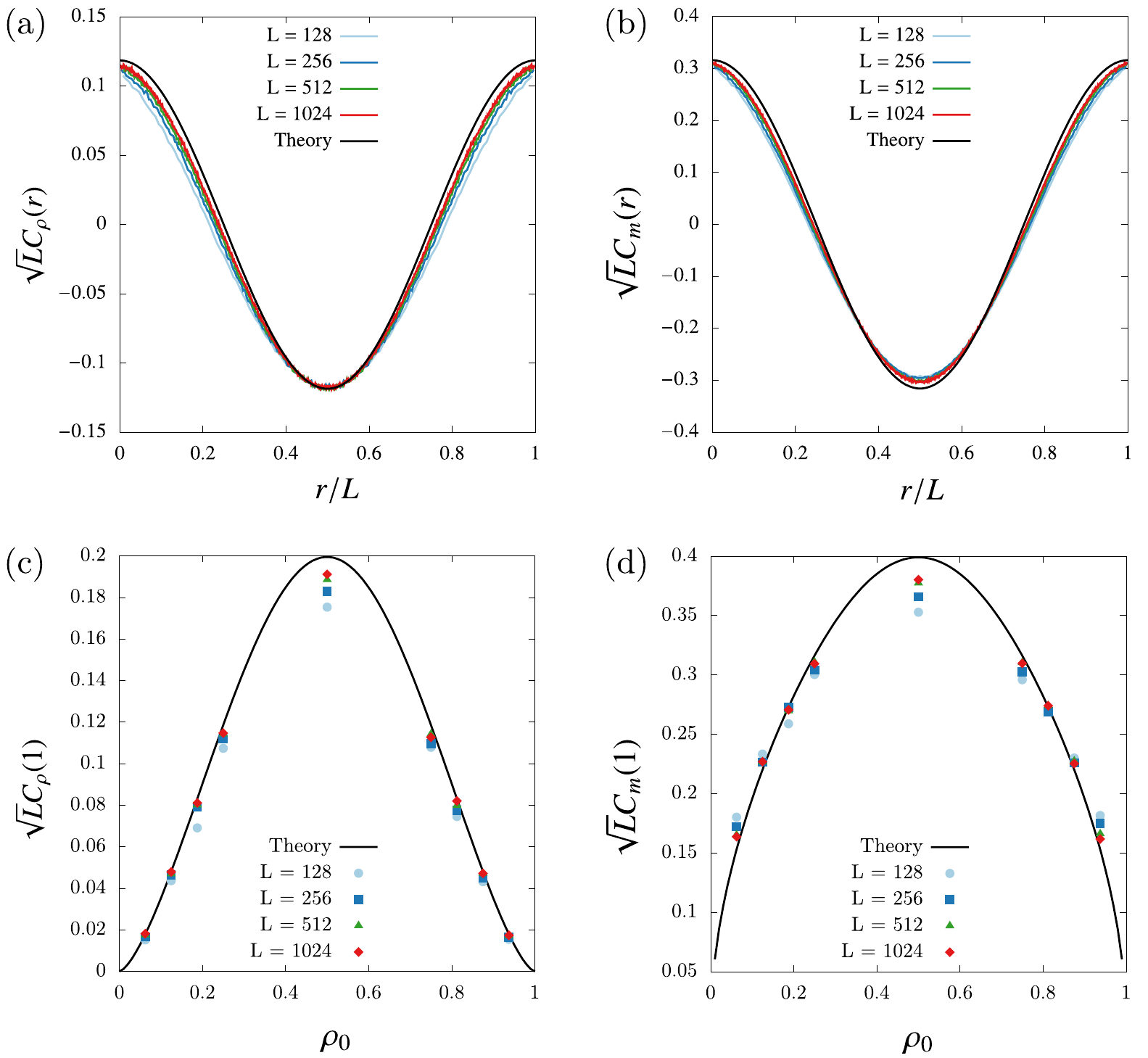}
\caption{(a) Connected density–density correlations of particles at a point on the critical surface of the scaled LH model. The separation between points scales with system size, while the correlation magnitude decays as $1/\sqrt{L}$. As the system size increases, the data approach the theoretical prediction in Eq.~\eqref{eq:scaled_critical_correlations}. (b) The corresponding tilt–tilt correlations likewise agree with the theoretical form for tilts in Eq.~\eqref{eq:scaled_critical_correlations}. Panels (a) and (b) are shown for a system with particle density fixed at $\rho_{0} = 0.25.$ (c) To examine Eq.~\eqref{eq:scaled_critical_correlations} as a function of $\rho_{0}$, we compare its predicted amplitude, $(\sqrt{L} C(0))$, with numerical simulation results. In the simulations, the on-site correlation $C(0)$ contains a short-range local contribution in addition to the regular long-range part described by the continuum theory, which complicates a direct comparison. However, since the theoretical form $C(r)\propto \cos(r/L)$ is essentially constant in the limit $r/L\to 0$, the nearest-neighbor correlation $C(1)$ serves as a good proxy for the regular amplitude. We therefore compare the theoretical $(\sqrt{L} C(0))$ with the simulated $(\sqrt{L} C(1))$. (d) Analogous plot for the tilt correlations. The parameters used are $\alpha = \pi/\sqrt{2}$ and $\beta = \beta^{\prime} = \pi\sqrt{2}$.}
\label{fig:critical_cor}
\end{figure}

We can infer the corresponding behavior in real space for the scaled model from the behavior of the modes. Because the dynamics are essentially governed by a single dominant mode, the formation of macroscopic clusters of fluctuating size does not appear to be possible in the scaled model. Moreover, the fact that the first mode decreases with increasing system size implies that the correlations will no longer be of $\mathcal{O}(1)$ magnitude; instead, their strength will diminish with system size. This is precisely what we find: Figure~\ref{fig:critical_cor} shows the two-point correlation function of the system at a critical point in the scaled model. We find that the correlation strength diminishes in proportion to the system size.

This type of critical behavior, controlled by a single mode, is already known from the sABC model. There, an anomalous long-range correlation function decaying as $1/\sqrt{L}$ was found. Since our results follow the same trend, we seek to provide an analytical description of this correlation function in the following section using the same method that was employed for the sABC model.
%%%%%%%%%%%%%%%%%%%%%%%%%%%%%%%%%%%%%%%%%%%%%%%%%%%%%%%%%%%%%%%%%%%%%%%

%%%%%%%%%%%%%%%%%%%%%%%%%%%%%%%%%%%%%%%%%%%%%%%%%%%%%%%%%%%%%%%%%%%%%%%5
%%%%%%%%%%%%%%%%%%%%%%%%%%%%%%%%%%%%%%%%%%%%%%%%%%%%%%%%%%%%%%%%%%%%%%%

%%%%%%%%%%%%%%%%%%%%%%%%%%%%%%%%%%%%%%%%%%%%%%%%%%%%%%%%%%%%%%%%%%%%%%%%%%%%%%%%%%%%%

\section{Critical behavior in the scaled Light-Heavy model}
\label{sec:Critical behavior in sLH}
%%%%%%%%%%%%%%%%%%%%%%%%%%%%%%%%%%%%%%%%%%%%%%%%%%%%%%%%%%%%%%%%%%%%%%%%%%%%%%%%%%%%%%
%%%%%%%%%%%%%%%%%%%%%%%%%%%%%%%%%%%%%%%%%%%%%%%%%%%%%%%%%%%%%%%%%%%%%%%5
%%%%%%%%%%%%%%%%%%%%%%%%%%%%%%%%%%%%%%%%%%%%%%%%%%%%%%%%%%%%%%%%%%%%%%%
\subsection{Two-point correlation function}
As established in the previous sections, the critical behavior of the scaled model is governed almost entirely by the first Fourier mode. It is therefore an excellent approximation to truncate the Fourier expansion of the local densities at the first mode as
\begin{eqnarray}
    \rho(x,t) = \rho_{0} + R_{1}(t)e^{i 2\pi x} + R^{*}_{1}(t)e^{-i 2\pi x}, \\
    m(x,t) = m_{0} + M_{1}(t)e^{i 2\pi x} + M^{*}_{1}(t)e^{-i 2\pi x}.
    \label{eq:density_in_Fourier_basis_truncated}
\end{eqnarray}
Here, $R_{n}$ is connected to $Q_{n}$ in Eq.~\eqref{eq:density_fm} through the straightforward relation $Q_{n} = \lvert R_{n} \rvert$.
The corresponding steady-state two-point correlation function can thus be expressed as
\begin{equation}
\begin{aligned}
\langle \rho(x)\rho(x')\rangle - \rho_0^2
&=
\Bigl\langle
\Bigl(
\rho_0
+ R_1^{\mathrm{ss}} e^{i2\pi x}
+ (R_1^{\mathrm{ss}})^* e^{-i2\pi x}
\Bigr) \\
&\qquad\qquad\times
\Bigl(
\rho_0
+ R_1^{\mathrm{ss}} e^{i2\pi x'}
+ (R_1^{\mathrm{ss}})^* e^{-i2\pi x'}
\Bigr)
\Bigr\rangle
- \rho_0^2.
\end{aligned}
\label{eq:cor_defn}
\end{equation}
for the heavy particles, and an analogous expression holds for the down tilts.

Evaluating the two-point correlations of the system at criticality therefore reduces to determining the long-time behavior of $R_{1}(t)$ and $M_{1}(t)$ and then substituting these results into Eq.~\eqref{eq:cor_defn} to obtain the correlation function. We will show that, in the steady state, the fluctuating hydrodynamic description of the system predicts that $R_{1}^{\text{ss}}, M_{1}^{\text{ss}} \sim L^{-1/4}$, which in turn leads to a connected correlation function scaling as $1/\sqrt{L}$. This behavior is consistent with the numerical results obtained in the previous section. A detailed derivation of the correlation function, following the corresponding procedure used in the sABC model~\cite{Gerschenfeld2012}, is presented below.

We begin with the expansion of the density fields in the Fourier basis:
\begin{align}
\rho(x,t) &= \rho_{0}
+ R_{1}(t)e^{i2\pi x} + R_{1}^{*}(t)e^{-i2\pi x}
+ R_{2}(t)e^{i4\pi x} + R_{2}^{*}(t)e^{-i4\pi x} \nonumber\\
&\quad + R_{3}(t)e^{i6\pi x} + R_{3}^{*}(t)e^{-i6\pi x}
+ \cdots , \nonumber\\[2mm]
m(x,t) &= m_{0}
+ M_{1}(t)e^{i2\pi x} + M_{1}^{*}(t)e^{-i2\pi x}
+ M_{2}(t)e^{i4\pi x} + M_{2}^{*}(t)e^{-i4\pi x} \nonumber\\
&\quad + M_{3}(t)e^{i6\pi x} + M_{3}^{*}(t)e^{-i6\pi x}
+ \cdots .
\label{eq:density_in_Fourier_basis_full}
\end{align}
%%%%%%%%%%%%%%%%%%%%%%%%%%%%%%%%%%%%%%%%%%%%%%%%%%%%%%%%%%%%%%%%%%%%%%%
Recall that, due to the net zero tilt condition imposed at the outset, the global density of down tilts satisfies $m_{0}=1/2$ for our system. We will substitute this value later on.
Starting from the fluctuating hydrodynamic description of the system given in Eq.~\eqref{hyd1}, the deterministic expressions for the currents associated with these density fields are
\begin{align}
j_{\rho}(x, t) &= -D\partial_x \rho(x,t)
+ 2\alpha\,\rho(x,t)\bigl[1-\rho(x,t)\bigr]\bigl[2m(x,t)-1\bigr],
\nonumber
\\
j_m(x,t) &= -D\partial_x m(x,t)
+ 2\beta\,m(x,t)\bigl[1-m(x,t)\bigr]\bigl[1-2\rho(x,t)\bigr],
\label{eq:current_gen_1}
\end{align}
when $\beta=\beta^{\prime}$. We adopt the assumption $\beta = \beta^{\prime}$ in order to simplify the analysis. As a result, the conclusions drawn from Eq.~\eqref{eq:current_gen_1} are valid only on this subspace and should be interpreted as representative of the behavior in the scaled limit, rather than as fully general. On this plane, we may write $\beta=\beta^{\prime}=c\alpha$, with $c$ a constant. Since our goal is to use Eq.~\eqref{eq:current_gen_1} to characterize the behavior of $R_{1}(t)$ and $M_{1}(t)$ at and close to the critical point, we can recast the system of equations as
\begin{align}
j_{\rho} &= -D\partial_x \rho(x,t)
+ 2\alpha_c (1+\gamma)\,\rho(x,t)\bigl[1-\rho(x,t)\bigr]\bigl[2m(x,t)-1\bigr],
\nonumber
\\
j_m &= -D\partial_x m(x,t)
+ 2\alpha_{c}(1+\gamma)\,c m(x,t)\bigl[1-m(x,t)\bigr]\bigl[1-2\rho(x,t)\bigr],
\label{eq:current_gen_2}
\end{align}
where $\alpha_{c}$ denotes the critical-point value predicted by the linear stability analysis in Eq.~\eqref{eq:lsa2}, which yields $\alpha_c = \pi D / \sqrt{c\rho_0(1-\rho_0)}$ upon imposing $\beta=\beta^{\prime}=c\alpha$. The parameter $\gamma \ll 1$ measures the deviation from criticality, with $\gamma > 0$ indicating the ordered phase.

Since both the total number of particles and tilts are conserved in our system, the currents $j_{\rho}(x,t)$ and $j_{m}(x,t)$ in Eq.~\eqref{eq:current_gen_2} must, of course, obey the continuity equation, which reads
\begin{equation}
\partial_{\bar{\tau}} \boldsymbol{\phi}
=
-\partial_x \mathbf{j},
\qquad
\boldsymbol{\phi}
=
\begin{pmatrix}
\rho\\
m
\end{pmatrix},
\quad
\mathbf{j}
=
\begin{pmatrix}
j_\rho\\
j_m
\end{pmatrix}.
\label{eq:continuity}
\end{equation}

Expanding Eq.~\eqref{eq:continuity} using Eq.~\eqref{eq:density_in_Fourier_basis_full}, we obtain
\begin{align}
j_{\rho}(x,t) &= j_{\rho0}
+ \frac{i}{2\pi}\dot{R}_{1}e^{i2\pi x} - \frac{i}{2\pi}\dot{R}_{1}^{*}e^{-i2\pi x}
+ \frac{i}{4\pi}\dot{R}_{2}e^{i4\pi x} - \frac{i}{4\pi}\dot{R}_{2}^{*}e^{-i4\pi x} \nonumber\\
&\quad + \frac{i}{6\pi}\dot{R}_{3}e^{i6\pi x} - \frac{i}{6\pi}\dot{R}_{3}^{*}e^{-i6\pi x}
+ \dots,
\label{eq:current_lhs}
\end{align}
and an analogous expression for the tilt field $m(x,t)$. By expanding Eq.~\eqref{eq:current_gen_2} in a similar manner, one can derive an alternative expression for the currents corresponding to the two density fields. Matching the two expressions after projecting them onto individual modes leads to equations of the form $\dot{R}_{i} = \mathcal{F}(\rho_{0}, R_{1}, R_{2}, \dots)$. Solving this system of equations in full generality would be extremely difficult were it not for the fact that the first mode becomes dominant at criticality. This allows us to neglect higher-order modes and obtain a much more manageable set of differential equations for $\dot{R_{1}}$ and $\dot{M_{1}}$. In doing so, one must carefully consider the relative magnitudes of the various terms to ensure that the resulting approximations for $R_{1}(\bar{\tau})$ and $M_{1}(\bar{\tau})$ are as accurate as possible.

The dominance of the first mode over higher modes implies $\rho_{0} \ll R_{1} \ll R_{2,3,4,\dots}$, and similarly for the tilt field. Furthermore, proximity to the critical point justifies assuming critical slowing down, which implies $\dot{R}_{1} \ll R_{1}$ (and similarly $\dot{M}_{1} \ll M_{1}$). This leads to two important consequences. First, when explicitly writing the equation for $\dot{R_{1}}$, the terms involving $\rho_{0}$ and $R_{1}, M_{1}$ cancel each other, leaving only a small correction term that contributes to $\dot{R}_{1}$. Second, the mode immediately adjacent to the first one, namely $R_{2}$, may influence $\dot{R}_{1}$, which motivates describing its dynamics relative to $R_{1}$. Both issues can be addressed by considering only the leading-order behavior of the projections onto the first and second modes, respectively.

By projecting Eqs.~\eqref{eq:current_gen_2} and \eqref{eq:continuity} onto the first Fourier mode and keeping only leading-order contributions, we obtain
\begin{equation}
    2 i D \pi R_{1}(t) = \frac{2\pi D}{\sqrt{c \rho_{0}(1-\rho_{0})}} \, 2\rho_{0} (1-\rho_{0})M_{1}(t) + \mathcal{O}(M_{1}^{2}(t)),
    \label{eq:leading_first_mode}
\end{equation}
which reduces to
\begin{equation}
M_{1}(t)
=
\left[
i\sqrt{
\frac{c}
     {4\rho_{0}(1-\rho_{0})}
}
\right]
R_{1}(t)
+
\epsilon,
\label{eq:m1_to_r1}
\end{equation}
where $\epsilon$ represents a correction term accounting for the omission of higher-order modes.

To determine the leading-order behavior of the second-mode amplitude relative to the first mode, we follow the same procedure by projecting onto the second mode. This yields
\begin{equation}
\begin{aligned}
i4\pi D R_{2}(t)
&=
\frac{2\pi D}{\sqrt{c\rho_{0}(1-\rho_{0})}}\Biggl[-2(2\rho_{0}-1)M_{1}(t)R_{1}(t) + 2\rho_{0}(1-\rho_{0})M_{2}(t)\Biggr],
\\[0.5ex]
i4\pi D M_{2}(t)
&=
\frac{2\pi D\sqrt{c}}{\sqrt{\rho_{0}(1-\rho_{0})}}\Biggl[(2\rho_{0}-1)M_{1}^{2}(t) - \frac{1}{2}R_{2}(t)\Biggr],
\end{aligned}
\label{eq:leading_second_mode}
\end{equation}
from which we can solve for
\begin{equation}
\begin{aligned}
R_{2}(t)
&=
\left[
\frac{(4-c)(1-2\rho_{0})}
     {6\rho_{0}(1-\rho_{0})}
\right]
R_{1}^2(t),
\\[0.5ex]
M_{2}(t)
&=
\left[
\frac{i(1-c)(1-2\rho_{0})}{6}
\sqrt{
\frac{c}
     {\rho_{0}^{3}(1-\rho_{0})^{3}}
}
\right]
R_{1}^2(t).
\end{aligned}
\label{eq:r2m2_to_r1}
\end{equation}
With these results available, one can analyze the next-to-leading-order behavior of the equations projected onto the first mode:
\begin{equation}
\begin{aligned}
\frac{i}{2\pi}\dot{R}_{1}(t)
&= i2\pi D\gamma R_{1}(t) \\
&\quad
+\frac{2\pi D}{\sqrt{c\rho_{0}(1-\rho_{0})}}
\Bigl[
2\rho_{0}(1-\rho_{0})\epsilon
-4|R_{1}(t)|^{2}M_{1}(t)
-2R_{1}^{2}(t)M_{1}^{*}(t)
\\
&\qquad\qquad\qquad
+2(1-2\rho_{0})R_{2}(t)M_{1}^{*}(t)
+2(1-2\rho_{0})M_{2}(t)R_{1}^{*}(t)
\Bigr],
\\[1.2ex]
\frac{i}{2\pi}\dot{M}_{1}(t)
&= i2\pi D\gamma M_{1}(t) \\
&\quad
+\frac{2\pi D}{\sqrt{c\rho_{0}(1-\rho_{0})}}
\Bigl[
-i\sqrt{c\rho_{0}(1-\rho_{0})}\,\epsilon
+4c\lvert M_{1}(t)\rvert^{2}R_{1}(t)
+2cM_{1}^{2}(t)R_{1}^{*}(t)
\\
&\qquad\qquad\qquad
-2c(1-2\rho_{0})M_{2}(t)M_{1}^{*}(t)
\Bigr].
\end{aligned}
\label{eq:next-to-leading_firstmode}
\end{equation}
However, the above equations contain the unknown correction term $\epsilon$. This term can be eliminated by taking the linear combination
\begin{equation}
  u\,\eqref{eq:next-to-leading_firstmode}\mathrm{(a)}
+ v\,\eqref{eq:next-to-leading_firstmode}\mathrm{(b)},  
\label{eq:lincombo}
\end{equation}
where
$u=i\sqrt{c\rho_{0}(1-\rho_{0})}$ and
$v=2\rho_{0}(1-\rho_{0})$.
This yields
\begin{align}
\partial_{t}X(t)
&=
4\pi^{2}D
\bigl(
\gamma
- f_{X}(c,\rho_{0})\lvert X(t)\rvert^{2}
\bigr)X(t),
\qquad
X = R_{1}, M_{1},
\\[1ex]
\label{eq:ndefn}
\mathcal{N}(c,\rho_{0})
&=
(2c^{2}-c+2)(2\rho_{0}-1)^{2}
+ 6(1+c)\rho_{0}(1-\rho_{0}),
\\[1ex]
f_{R_{1}}(c,\rho_{0})
&=
\frac{\mathcal{N}(c,\rho_{0})}
{12\,\rho_{0}^{2}(1-\rho_{0})^{2}},
\\
f_{M_{1}}(c,\rho_{0})
&=
\frac{\mathcal{N}(c,\rho_{0})}
{3c\,\rho_{0}(1-\rho_{0})}.
\end{align}
Up to this point, we have considered only the deterministic current specified in Eq.~\eqref{eq:current_gen_2}. However, in finite systems there is an extra contribution originating from the noise fields, as indicated by Eq.~\eqref{hyd1}, which describes the evolution of the density fields. We now add the noise term from Eq.~\eqref{eq:hyd2} to our expressions for the current and expand it using Eq.~\eqref{eq:density_in_Fourier_basis_full}. The leading-order contribution of the projection of these current expressions onto the first mode is not affected by the corresponding projected noise term, which scales as $\rho_{0}/\sqrt{L}$. Consequently, Eq.~\eqref{eq:leading_first_mode} remain valid even after including the noise fields $\eta_{\rho}(x,t)$ and $\eta_{m}(x,t)$.

The situation is different for the next-to-leading-order behavior of the projected current equations in the first mode, given in Eq.~\eqref{eq:next-to-leading_firstmode}, where the relevant terms are of much smaller order and must therefore be corrected by adding the term $\chi_{\rho,m}^{(1)}(t) = \int_{0}^{1}dx\, e^{-i2\pi x}\eta_{\rho,m}(x,t)$. In contrast to the leading-order terms in the first-mode current projections, the leading-terms in the expressions projected onto the second mode are smaller in magnitude and hence require explicit inclusion of the noise contribution $\chi_{\rho,m}^{(2)}(t) = \int_{0}^{1}dx\, e^{-i4\pi x}\eta_{\rho,m}(x,t)$, which modifies Eq.~\eqref{eq:leading_second_mode}. 

However, our use of Eq.~\eqref{eq:leading_second_mode} is restricted to determining $R_{2}(t)$ and $M_{2}(t)$, which are then substituted into Eq.~\eqref{eq:next-to-leading_firstmode} to solve for $R_{1}(t)$. In Eq.~\eqref{eq:next-to-leading_firstmode}, the quantities $R_{2}$ and $M_{2}$ appear only in products with $R_{1}^{*}$ or $M_{1}^{*}$. Consequently, the $\chi_{\rho,m}^{(2)}(x,t)$ term entering the modified form of Eq.~\eqref{eq:next-to-leading_firstmode} is much smaller than the already present $\chi_{\rho,m}^{(1)}(x,t)$ contribution and can safely be neglected. Therefore, we do not actually need to modify Eq.~\eqref{eq:leading_second_mode} to incorporate noise effects. 

In summary, to account for the noise terms that appear in the expressions of the current for finite systems, it suffices to insert the appropriate noise term, $\chi_{\rho,m}^{(1)}(x,t)$, into Eq.~\eqref{eq:next-to-leading_firstmode}, with its correlator $\langle\chi_{\rho,m}^{(1)}(\bar{\tau})\chi_{\rho,m}^{*~(1)}(\bar{\tau}^{\prime})\rangle$ given by Eq.~\eqref{eq:noise_correlations}, while $\langle\chi_{\rho,m}^{(1)}(\bar{\tau})\chi_{\rho,m}^{(1)}(\bar{\tau}^{\prime})\rangle = 0$. Applying the same linear combination specified in Eq.~\eqref{eq:lincombo} to the noise terms $\chi_{\rho,m}(\bar{\tau})$, as was previously carried out for the deterministic expression in the analysis, then leads to
\begin{equation}
    \partial_{t} X(t) = 
 4\pi^{2} D \Biggl(\gamma - f_{X}(c,\rho_{0}) \lvert X(t)\rvert^{2}\Biggr) X(t) + \frac{\mu_{X}(t)}{L},
 \label{eq:r1_equation}
\end{equation}
where the correlations of the fields $\mu(x,t)$ are given by
\begin{align}
\langle \mu_{R_{1}}(t)\mu_{R_{1}}^{*}(t^{\prime})\rangle
&=
2\pi^{2}D\frac{1+c}{c}\rho_{0}(1-\rho_{0})\,
\delta(t-t'),
\nonumber\\[1ex]
\langle \mu_{M_{1}}(t)\mu_{M_{1}}^{*}(t^{\prime})\rangle
&=
\frac{\pi^{2}D}{2}(1+c)\,
\delta(t-t').
\label{eq:mu_correlations}
\end{align}
%%%%%%%%%%%%%%
Eq.~\eqref{eq:r1_equation} can be cast into a much simpler form by  introducing the rescaled variables
\begin{align}
\label{eq:g_defn}
R_{1}(t)
&=
h_{R_1}(c,\rho_0)
g(\bar{\tau}),
\nonumber\\[1ex]
M_{1}(t)
&=
h_{M_1}(c,\rho_0)
g(\bar{\tau}),
\end{align}
along with
\begin{align}
\label{eq:rescaled_variables}
\bar{\tau} 
&=
\pi^{2}D
\sqrt{
\frac{2\mathcal{N}(c,\rho_{0})}{3\rho_0(1-\rho_0)}
\frac{1+c}{c}}
\,\frac{t}{\sqrt L},
\\[1ex]
\bar{\gamma}
&=
\sqrt{
\frac{24\rho_0(1-\rho_0)}
{\mathcal{N}(c,\rho_{0})}
\frac{c}{1+c}}
\,\sqrt {L}~\gamma,
\end{align}
where,
\begin{align}
\label{eq_h_def1}
h_{R_1}(c,\rho_0)
&=
\sqrt[4]{
\frac{
6(1+c)\rho_0^3(1-\rho_0)^3
}{
c\,\mathcal N(c,\rho_0)L
}},
\\
\label{eq_h_def2}
h_{M_1}(c,\rho_0)
&=
\sqrt[4]{
\frac{
3c(1+c)\rho_0(1-\rho_0)
}{
8\,\mathcal N(c,\rho_0)L
}},
\end{align}
with $\mathcal{N}(c,\rho_0)$ defined in Eq~\eqref{eq:ndefn}.
Under these transformations Eq.~\eqref{eq:r1_equation} reduces to the overdamped Langevin equation
\begin{equation}
\partial_{\bar{\tau}}g(\bar{\tau})
=
\left(\bar{\gamma}-|g(\bar{\tau})|^{2}\right)g(\bar{\tau})
+\mu(\bar{\tau}),
\label{eq:rescaled_timeevltn}
\end{equation}
where
\[
\langle\mu(\bar{\tau})\mu^{*}(\bar{\tau}^{\prime})\rangle
=\delta(\bar{\tau}-\bar{\tau}^{\prime}).
\]
Eq.~\eqref{eq:rescaled_timeevltn} can be rewritten as a Fokker–Planck equation governing the dynamics of the rescaled variable \(g(\bar{\tau})\) in the complex plane. The probability density \(P(g_x,g_y,\bar{\tau})\) then satisfies
\begin{equation}
    \partial_{\bar{\tau}}P
    =
    \frac{1}{r}\partial_{r}\biggl[(r^{2}-\bar{\gamma})r^{2}P + \frac{1}{4}r\partial_{r}P\biggr]
    + \frac{1}{4 r^{2}}\partial_{\theta}^{2}P,
    \label{eq:fp}
\end{equation}
where \(\mathbf{r}=(g_{x},g_{y})\), with $r = \sqrt{g_{x}^2 + g_{y}^2}$ and $\theta=\arctan{\left(\frac{g_y}{g_x}\right)}$, closely following what was observed in the sABC model~\cite{Gerschenfeld2011}. The corresponding stationary solution of Eq.~\eqref{eq:fp} is rotationally invariant in the complex plane and reads
\begin{equation}
    P_{\text{ss}}(\mathbf{r},\bar{\tau})= e^{2\bar{\gamma}r^{2}-r^{4}}.
\end{equation}
As a consequence, we obtain
\begin{equation}
\langle R_1\rangle
=
\langle R_1^*\rangle
=
\langle R_1^2\rangle
=
\langle (R_1^*)^2\rangle
=0,
\end{equation}
and the same relations hold for \(M_1\). The connected two-point functions therefore simplify to
\begin{equation}
\langle\rho(x)\rho(x')\rangle-\rho_0^2
=
2\langle|R_1(t_{\infty})|^2\rangle
\cos\bigl(2\pi (x-x')\bigr),
\end{equation}
with an analogous form for \(m(x)\), where
\begin{equation}
    \langle \lvert X(t_{\infty})\rvert^{2}\rangle
    = h_{X}^{2}(c,\rho_{0})\,
      \langle \lvert g(\bar{\tau}_{\infty})\rvert^{2}\rangle.
\end{equation}
Evaluating $\langle \lvert g(t_{\infty})\rvert^{2}\rangle$~\cite{Gerschenfeld2012}], we find
\begin{equation}
    \langle \lvert g(t_{\infty})\rvert^{2}\rangle
    =
    \frac{\int_{0}^{\infty} r^{3} e^{2\bar{\gamma}r^{2}-r^{4}}\,dr}
         {\int_{0}^{\infty} r\, e^{2\bar{\gamma}r^{2}-r^{4}}\,dr}
    = \bar{\gamma}
      +\frac{1}{2}\frac{e^{-\bar{\gamma}^{2}}}
                     {\int_{-\infty}^{\bar{\gamma}}e^{-z^{2}}dz}
    \equiv l(\bar{\gamma}),
\end{equation}
which, at the critical point \(\bar{\gamma}=0\), yields for the correlation functions
\begin{align}
\langle\rho(x)\rho(x')\rangle-\rho_0^2
&=
\frac{2}{\sqrt{\pi}}
\,h_{R_1}^{2}(c,\rho_0)
\cos\bigl(2\pi (x-x^{\prime})\bigr),
\nonumber
\\
\langle m(x)m(x')\rangle-m_0^2
&=
\frac{2}{\sqrt{\pi}}
\,h_{M_1}^{2}(c,\rho_0)
\cos\bigl(2\pi (x-x^{\prime})\bigr),
\label{eq:scaled_critical_correlations}
\end{align}
%%%%%%%%%%%%%%
where $h_{R_1}$ and $h_{M_1}$ are defined in Eqs.~\eqref{eq_h_def1} and \eqref{eq_h_def2}. Thus, in the scaled model, the correlations are dominated by the first Fourier mode and consequently acquire a cosine-like form. This is in stark contrast to the cusp-like correlations characteristic of FDPO. In FDPO, the cusp exponent can be estimated using the Independent Interval Approximation (IIA), when the cluster size distribution follows a power-law form. We now examine how this picture changes in the scaled model by studying its cluster size distribution and applying the IIA framework.

\subsection{Cluster size distribution}
\label{subsection:cluster size distribution}
%%%%%%%%%%%%%%%%%%%%%%%%%%%%%%%%%%%%%%%%%%%%%%%%%%%%%%%%%%%%%%%%%%%%%%%5
\begin{figure}[t!]
\centering
\includegraphics[width=1.0\linewidth] {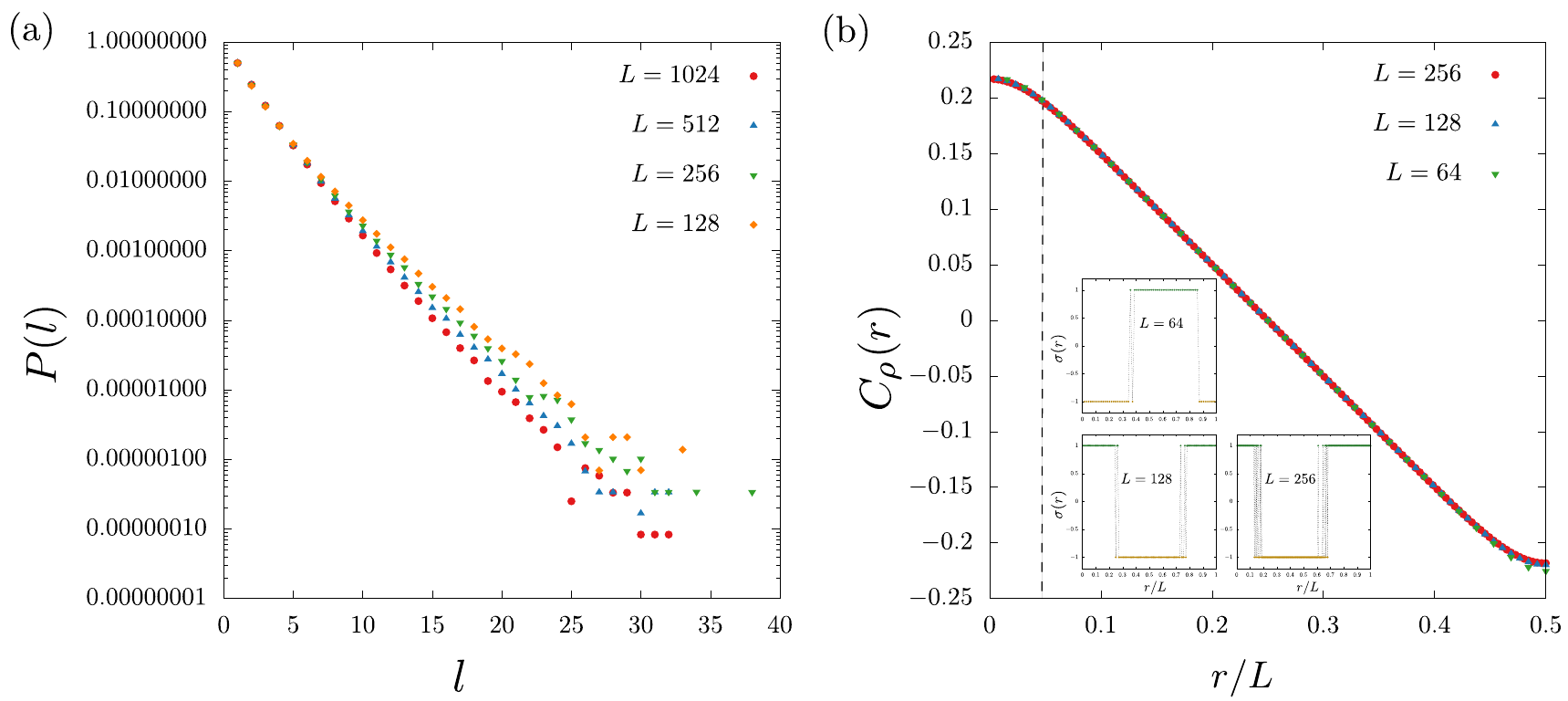}
\caption{(a) At criticality, the cluster size distribution in the scaled LH model follows an exponential form. The parameters used are $\alpha =\pi/\sqrt{2},~\beta = \beta^{\prime}=\pi \sqrt{2}$. (b) The connected density-density correlation function of particles in the phase-separated regime decays linearly, in agreement with Porod’s law, after an initial plateau region that scales with $L$, which arises due to the presence of broad interfaces. The insets display the corresponding spin configuration of the particles, highlighting the broadened interface. The parameters used are $\alpha =\beta = \beta^{\prime}= 15.$} 
\label{fig:porod_law}
\end{figure}

%%%%%%%%%%%%%%%%%%%%%%%%%%%%%%%%%%%%%%%%%%%%%%%%%%%%%%%%%%%%%%%%%%%%%%%
The scaled LH model exhibits an exponential cluster size distribution along the critical surface (see Fig.~\ref{fig:porod_law}~a). This differs fundamentally from the power-law distribution observed in FDPO, where the IIA can be used to compute the exponent appearing in the correlation function~\cite{Das2000,Barma2019}. Applying the same approach here, using the numerically observed distribution
\[
P(l)\sim \frac{1}{\zeta}e^{-l/\zeta},
\]
a naive IIA calculation would predict correlations of the form
\[
C(r)\sim e^{-2r/\zeta}.
\]
However, this prediction disagrees with both our analytical results and simulations. The failure arises because the assumptions underlying the IIA break down in the scaled model: cluster sizes can no longer be treated as independent, uncorrelated random intervals, as they are in the unscaled model.

%%%%%%%%%%%%%%%%%%%%%%%%%%%%%%%%%%%%%%%%%%%%%%%%%%%%%%%%%%%%%%%%%%%%%%%
%%%%%%%%%%%%%%%%%%%%%%%%%%%%%%%%%%%%%%%%%%%%%%%%%%%%%%%%%%%%%%%%%%%%%%%
\section{Ordered phase and Porod law}
\label{sec:ordered_phase_porod}
%%%%%%%%%%%%%%%%%%%%%%%%%%%%%%%%%%%%%%%%%%%%%%%%%%%%%%%%%%%%%%%%%%%%%%%
%%%%%%%%%%%%%%%%%%%%%%%%%%%%%%%%%%%%%%%%%%%%%%%%%%%%%%%%%%%%%%%%%%%%%%%
One of the most striking features of FDPO is that, although it is considered a critical state, it nevertheless exhibits several properties in common with standard phase-ordering systems. During coarsening, the characteristic length scale increases as $t^{1/z}$, and in the steady state it is determined by the system size $L$. In contrast to conventional phase ordering, however, FDPO violates Porod's law: rather than showing a linear decay of $C(r)$ for $\xi\ll r\ll L$, the correlation function develops a cusp singularity. This immediately raises the question of whether the breakdown of Porod's law is specific to FDPO, or if it also appears in the truly phase-separated states of the same model. In this section, we address this issue by analyzing the correlation function within the phase-separated regimes.

In the case of the strong phase separation seen in the unscaled model, compliance with Porod's law is fairly straightforward to anticipate. This is less evident, however, in the phase-separated regime of the scaled model, where one would instead expect relatively broad interfaces. Nevertheless, the system does obey Porod's law (see Fig.~\ref{fig:porod_law}~b), but only outside a core region, marked by the dashed line in the correlation plot, beyond which the correlation function exhibits a linear decay. We have also examined the coarse-grained density profiles in real space to support this interpretation; this is further suggested by the real-space configurations (shown in the inset of Fig.~\ref{fig:porod_law}~b), which indicate that the interfacial region itself scales with $L$. Such an $L$-dependent interfacial width can be naturally understood using the hydrodynamic description of the model given in Eq.~\eqref{eq:hyd2}.
%%%%%%%%%%%%%%%%%%%%%%%%%%%%%%%%%%%%%%%%%%%%%%%%%%%%%%%%%%%%%%%%%%%%%%%%%%%%%%%%%%%%%
\section{Discussion}
\label{sec:discussion}     
%%%%%%%%%%%%%%%%%%%%%%%%%%%%%%%%%%%%%%%%%%%%%%%%%%%%%%%%%%%%%%%%%%%%%%%%%%%%%%%%%%%%%
In this work, we investigated the critical and near-critical behavior of the Light-Heavy (LH) model. In the unscaled LH (uLH) model, the critical behavior is governed by Fluctuation-Dominated Phase Ordering (FDPO), which is characterized by macroscopic long-range correlations. We demonstrated that the homogeneous regime in the vicinity of this critical surface retains distinct local signatures of FDPO. Specifically, well-defined local domains emerge, governed by a correlation length that depends directly on the distance from the critical surface. Within this length scale, the system effectively mimics FDPO behavior, remaining insensitive to the fact that the system is globally in the off-critical homogeneous phase. This spatial insensitivity closely mirrors the role of the correlation length in conventional critical phenomena. To formalize this local FDPO phenomenon, we introduced a unified scaling function that successfully interpolates the two-point correlation functions across both the critical and off-critical regimes.

We subsequently extended our analysis to the scaled limit, where the drive parameter scales as $1/L$. This weakly driven regime yields an effectively diffusive macroscopic limit, which facilitates rigorous analytical treatment while still maintaining a remarkably rich near-equilibrium phase diagram. Focusing on the critical phase of this scaled model, we conclusively identified it as non-FDPO. Unlike the multi-mode structure typical of FDPO in the unscaled limit, a single long-wavelength Fourier mode controls the critical fluctuations, drawing strong parallels to the behavior observed in the scaled ABC (sABC) model. While this phase preserves the system size scaling of the separation seen in FDPO two-point correlations, its amplitude vanishes in the thermodynamic limit, scaling as $L^{-1/2}$. This single-mode dominance drastically simplifies the underlying dynamics, enabling us to derive an exact analytical description of the first Fourier mode in the steady state. This derivation directly yields the functional form of the two-point correlation function, which we confirmed to be in excellent agreement with extensive Monte Carlo simulations. These contrasting behaviors between the unscaled and scaled limits highlight the delicate balance between diffusive relaxation and driven transport, offering a broader perspective on fluctuation-dominated ordering in nonequilibrium systems. 
%%%%%%%%%%%%%%%%%%%%%%%%%%%%%%%%%%%%%%%%%%%%%%%%%%%%%%%%%%%%%%%%%%%%%%%%%%%%%%%%%%%%%

\section{Acknowledgements}
We thank R. Bhutia, S. Chakraborty and S. Sircar for useful discussions. M.B acknowledges the support of the Indian National Science Academy. This project was funded by intra-mural funds at TIFR Hyderabad from the Department
of Atomic Energy (DAE), Government of India, under Project Identification No. RTI4007.

%%%%%%%%%%%%%%%%%%%%%%%%%%%%%%%%%%%%%%%%%%%%%%%%%%%%%%%%%%%%%%%%
\bibliographystyle{spphys}
\bibliography{lh}

%\addcontentsline{toc}{section}{\protect\bibname}
%\begin{thebibliography}{10}

%\end{thebibliography}
\end{document}